\documentclass[superscriptaddress,showkeys,showpacs]{revtex4}
\usepackage{feynmf}
\usepackage{graphicx}
\usepackage[tbtags]{amsmath}
\input{epsf.sty}
\allowdisplaybreaks[4]
\begin{document}
\begin{fmffile}{fmfp}
\title{Next to leading order evolution of SIDIS processes in the forward region}
\thanks{Partially supported by CONICET, Fundaci\'on Antorchas, UBACYT and 
ANPCyT, Argentina.}

\author{A. Daleo}
\email{daleo@fisica.unlp.edu.ar}
\affiliation{Laboratorio de F\'{\i}sica Te\'{o}rica\\ Departamento de
F\'{\i}sica, Facultad de Ciencias Exactas\\
Universidad Nacional de La Plata\\ C.C. 67 - 1900 La
Plata,  Argentina}
\author{R. Sassot}
\affiliation{Departamento de F\'{\i}sica,
Universidad de Buenos Aires\\ Ciudad Universitaria, Pab.1 (1428)
Buenos Aires, Argentina}

\begin{abstract}
We compute the order $\alpha_s^2$ quark initiated corrections to semi-inclusive
deep inelastic scattering extending the approach developed recently for the gluon
contributions. With these corrections we complete the order $\alpha_s^2$ QCD 
description of these processes, verifying explicitly the factorization of 
collinear singularities. We also obtain the corresponding NLO evolution kernels,
relevant for the scale dependence of fracture functions. 
We compare the non-homogeneous evolution effects driven by these kernels with 
those obtained at leading order accuracy and discuss their phenomenological 
implications.
\end{abstract}


\maketitle

\section*{Introduction}

Fracture functions play a fundamental role in the description of one particle
inclusive deep inelastic scattering \cite{ven,grau,massi,npb}. They have been 
incorporated into the analysis of a variety of specific processes, 
although only
 up to leading order accuracy \cite{ppp,prd98,dss,ds,prd97}, since their scale 
dependence, which includes non homogeneous effects, had not been studied at 
higher orders until recently. At the lowest order in QCD perturbation (LO), 
these functions account for processes in which the final state particle is
produced in the direction of the target remnant, and can be interpreted as
the probability to find a parton in an already fragmented nucleon target 
\cite{ven}.
Their non homogeneous scale dependence is found 
to be strongly suppressed at LO in most of the kinematic domain, particularly 
in the kinematic region covered by the present experiments, and thus neglected.
However, there are no indications if this suppression is also present at NLO 
level.

In reference \cite{oad} the gluon initiated 
corrections to one particle inclusive deep inelastic scattering were computed 
up to order $\alpha_s^2$ for the first time. There it was shown that at 
variance with the fully inclusive case, in the one particle inclusive cross
section at this order, collinear configurations lead to singular distributions
 in more than one variable which can not be handled with the standard 
procedure. 
However a suitable method for the 
prescription of the characteristic entangled singularities was developed,
verifying explicitly the factorization of collinear singularities and 
obtaining the corresponding non homogeneous evolution kernels.
Regarding the phenomenological consequences of these corrections, it was 
found that the non homogeneous contributions to the evolution 
equations at NLO may be noticeably larger than the ones found at LO.
Of course, gluon initiated corrections do not provide all the kernels 
relevant for the scale dependence of fracture functions, a consistent 
assessment of this dependence requires also the consideration of quark 
initiated processes at the same order.   

In the present paper we extend the computation of the ${\cal O}(\alpha_s^2)$ 
corrections to one particle inclusive DIS for the case of quarks in the 
initial state. These last processes involve two caveats, a more complex 
singularity structure in the partonic cross sections due to the presence of
virtual corrections, and complications associated to the flavor structure and
decomposition for these corrections.  

Quark initiated corrections imply the computation of several real and
virtual contributions which lead to a much more complicated singularity 
structure than in the gluon initiated case. The entanglement of 
singularities is precisely the most difficult point in the computation of
 higher order corrections to one particle inclusive processes. Nevertheless, 
we show that they can be handled extending the strategy implemented for 
the latter case. 

Regarding the flavor structure,  the explicit factorization of collinear 
singularities in the one particle inclusive cross section does not allow
to obtain all the evolution kernels individually, only certain 
combinations of them can be extracted. Consequently, at this point it is not
possible to achieve a complete flavor discrimination in the evolution of
fracture functions. However it is possible to define certain singlet and non 
singlet flavor combinations, identify their respective evolution kernels, and 
obtain the evolution of the cross section.

As in the case of gluon initiated corrections, NLO contributions to the 
non homogeneous evolution initiated by quarks are found to be comparable 
or even larger than the LO ones in some kinematical regions. 
In order to assess the significance of non homogeneous contributions relative
to homogeneous ones in the scale dependence, we implement
the full evolution equations for $M_{q,\pi^+/P}$ using as input a model 
estimate for these fracture functions at an initial scale.
The results we found indicate large non homogeneous effects at small $x_L$ and
$x_{B}$ but also suggest suppression for both LO and NLO contributions 
otherwise. On the other side, this estimate does not take into account the 
NLO non homogeneous kernels for gluon fracture functions, which show up at 
order $\alpha_s^3$ in the cross section. These kernels contribute indirectly 
to $M_q$ through the homogeneous terms in the evolution but have not been 
computed yet. 

In the following section we summarize the relevant kinematics and details 
about the phase space integration for the O($\alpha_s^2$) contributions to
the cross section, together with the conventions and notation adopted.
In section II we compute the corresponding real and virtual amplitudes 
associated to each of the quark initiated processes, we discuss the nature
of the singularities that contribute to them and the prescriptions required.
Then, in section III  we analyse the factorization of collinear singularities, 
and issues related to the flavor decomposition. There, we also present the 
evolution kernels obtained and derive the evolution equations.
Section IV discuss the phenomenological implications of the new corrections 
and finally we end with our conclusions.


\section{Notation and Kinematics}
We begin by reviewing the main features of the QCD description of one particle 
inclusive deep inelastic scattering processes. We consider the process 
\begin{equation}
l(l)+P(P)\longrightarrow l^{\prime}(l^\prime)+h(P_h)+X,
\end{equation}
where a lepton 
of momentum $l$ scatters off a nucleon of momentum $P$ with 
a lepton of momentum $l^{\prime}$ and a 
hadron $h$ of momentum $P_h$ tagged in the final state. The cross section 
for this process can be written as \cite{grau}
\begin{eqnarray}\label{eq:hadcsec}
\frac{d\sigma}{dx_B\, dy\, dv_h\, dw_h}=&&\sum_{i,j=q,\bar{q},g}\int_{x_B}^{1}
	\frac{du}{u}\int_{v_h}^{1} \frac{dv_j}{v_j}\int_{0}^{1}dw_j\,
	f_{i/P}\bigg(\frac{x_B}{u}\bigg)\,D_{h/j}\bigg(\frac{v_h}{v_j}\bigg)
	\,\frac{d\hat{\sigma}_{ij}}{dx_B\, dy\, dv_j\, dw_j}\,
	\delta(w_h-w_j)\nonumber\\
	&&+\sum_{i}\int_{\frac{x_B}{1-(1-x_B)v_h}}^{1}
	\frac{du}{u}\,\,M_{i,h/P}\left(\frac{x_B}{u},(1-x_B)v_h\right)
	\,(1-x_B)
	\,\frac{d\hat{\sigma}_{i}}{dx_B\, dy}\,\delta(1-w_h)
	\,,
\end{eqnarray}
where the sum is over all parton species. The leptonic final state, in the 
one photon exchange approximation, is described by the usual deep inelastic
 variables
\begin{equation}
Q^2=-q^2=-(l^{\prime}-l)^2\,,\;\,\,\,\,\,\,x_{B}=\frac{Q^2}{2 P\cdot q}\,,
\;\,\,\,\,\,\,
y=\frac{P\cdot q}{P\cdot l}\,,\;\,\,\,\,\,\,S_H=(P+l)^2\,,
\end{equation}
whereas, to describe the hadronic final state, we define
\begin{equation}
v_{h}=\frac{E_{h}}{E_{0}(1-x_{B})}\,,\:\:\,\,\,\,\,\,\,\,\,\,\,\,\,\,
	w_{h}=\frac{1-\cos \theta_{h}}{2}\,.
\end{equation}
$E_{h}$ and $E_{0}$ in the preceding formulae are the energies of the final 
state hadron and of the incoming proton in the $\vec{P}+\vec{q}=0$ frame, respectively. $\theta_{h}$ is the angle between the momenta of the hadron and that 
of the virtual photon in the same frame. The convolution variables $v_{i}$ and
$w_{i}$ in equation (\ref{eq:hadcsec}) are the partonic analogs of $v_{h}$ and 
$w_{h}$ and $u$ is related to the fraction of momentum of the incoming
parton $\xi$ by $\xi=x_B/u$.

The two terms in the r.h.s. of equation (\ref{eq:hadcsec}) stand for 
`current' and `target'  fragmentation processes respectively. 
In the first one  
$d\hat{\sigma}_{ij}$ represents the partonic 
cross section for the process $l+i\rightarrow l^{\prime}+j+X$, whereas
$f_{i/P}$ and $D_{h/j}$  are the usual partonic densities
and fragmentation functions. In the second term the
cross sections for the inclusive process $l+i\rightarrow l^{\prime}+X$,
$d\hat{\sigma}_{i}$, are convoluted with the fracture functions  $M_{i,h/P}$ 
which give the probability of finding parton $i$ in an already fragmented
target. In this last case, the hadron is produced in the direction of the
incoming proton (notice the $\delta(1-w_h)$ in the last factor of the 
second term). 

The partonic cross sections 
$d\hat{\sigma}_{i}$ and $d\hat{\sigma}_{ij}$ can be computed order by 
order in perturbation theory from the corresponding
parton-photon squared matrix elements, $i+\gamma\rightarrow X$ and 
$i+\gamma\rightarrow j+X$  respectively. 
The totally inclusive cross sections up to order $\alpha_s^2$, are well known
and can be found in reference \cite{zijli}. 
The lowest order one-particle inclusive cross section  (zeroth-order in $\alpha_s$) is given by
\begin{equation}
\frac{d\hat{\sigma}^{(0)}_{qq,M}}{dx_B\,dy\,dv\,dw}=
\frac{d\hat{\sigma}^{(0)}_{\bar{q}\bar{q},M}}{dx_B\,dy\,dv\,dw}=
\frac{ 4\pi\,\alpha^2\,e_q^2}{2\,x_B\,S_H}\,(2+\epsilon)\,Y_M\,
	\delta(1-u)\delta(1-v)\delta(w)\,, 
\end{equation}
where the outgoing quark is always produced backwards ($w=0$). 
The index $M$ stands for metric, longitudinal contributions 
are absent at tree level, and $Y_M=(1+(1-y)^2)/2y^2$. The result correspond to $d=4+\epsilon$ space-time 
dimensions.

The order-$\alpha_s$ corrections to the one-particle inclusive cross 
sections are also known. Expressions for the singular and finite terms 
computed within the framework of dimensional regularization\cite{Bol} 
can be found in \cite{grau} for the unpolarized scattering,
and in \cite{npb} for the polarized case. Finite contributions up to order 
$\epsilon$, needed for the factorization of collinear singularities at 
${\cal O}(\alpha_s^2)$, are given in \cite{oad}.

At order-$\alpha_s$ the fragmenting parton can be produced in any 
direction, including the forward one ($w=1$), but all singular contributions 
come either from the backward or from the forward direction. The former 
terms are factorized into partonic densities and fragmentation functions, 
whereas the forward singularities can only be factorized in the redefinition 
of fracture functions. This factorization gives rise, as we have already 
mentioned, to non-homogeneous terms in the evolution equations of fracture 
functions \cite{ven}.

At order-$\alpha_s^2$ the partonic cross sections develop singularities
in every direction in space. They receive contributions from both virtual 
and real amplitudes whose matrix elements have to be integrated over the 
phase space of the unobserved partons (i.e. those partons that are not 
attached to the fragmentation function). At this order, real contributions 
have three particles in the final state, and the corresponding phase space 
can be written as:
\begin{eqnarray}\label{eq:PS3}
dPS^{(3)}&=&
	\frac{Q^2}{(4\pi)^4}\,\frac{1}{\Gamma\left(1+\epsilon\right)}
	\left(\frac{Q^2}{4\pi}\right)^{\epsilon}\,
	\left(\frac{1-x_B}{x_B}\right)^{1+\epsilon/2}
	\left(\frac{u-x_B}{x_B}\right)^{\epsilon/2}\,
	v^{1+3\,\epsilon/2}
	\theta(w_r-w)\,
	(w_r-w)^{\epsilon/2}\, w^{\epsilon/2}\nonumber\\
	&&\times(1-w)^{\epsilon/2}
	dv\,dw\,\sin^{1+\epsilon}\beta_{1}\,\sin^{\epsilon}\beta_{2}\,
	d\beta_{1}\,d\beta_{2}\,.
\end{eqnarray}
The angles $\beta_{1}$ and $\beta_{2}$ are the polar and azimuthal angle
of the pair of unobserved partons in their center of mass frame. 
In order to enforce the kinematical constraints and deal with the 
singularities in the prescription stage, it is convenient to break up 
the integration range into three regions, $R=\mbox{B0}\cup \mbox{B1}\cup \mbox{B2}$
with  
\begin{eqnarray}\label{eq:regNLO}
&&\mbox{B0}=\{u\in[x_B,x_u],\:v\in[v_h,a],\:w\in[0,1]\}\nonumber\\
&&\mbox{B1}=\{u\in[x_B,x_u],\:v\in[a,1],\:w\in[0,w_r]\}\nonumber\\
&&\mbox{B2}=\{u\in[x_u,1],\:v\in[v_h,1],\:w\in[0,w_r]\}\,,
\end{eqnarray}   
where 
\begin{equation}\label{eq:aywr}
x_u=\frac{x_B}{(x_B+(1-x_B)v_h)}\,,\:\:\:
	a=\frac{(1-u)\,x_B}{u\,(1-x_B)}\,,\:\:\:
	w_r=\frac{(1-v)\,(1-u)\,x_B}{v\,(u-x_B)}\,.
\end{equation}
B1 and B2 receive contributions from both ${\cal O}(\alpha_s)$ and ${\cal O}(\alpha_s^2)$ processes, while B0 has only ${\cal O}(\alpha_s^2)$.  
Virtual contributions can have either one or two particles in the final state.
As it will be discussed in the next section, only diagrams with two particles 
in the final state give rise to contributions in the forward direction. In this
case, the phase space can be conveniently cast into
\begin{eqnarray}\label{eq:PS2}
dPS^{(2)}&=&\frac{1}{8\pi\,\Gamma\left(1+\epsilon/2\right)}
	\left(\frac{Q^2}{4\pi}\right)^{\epsilon/2}
	\frac{u\,(1-x_B)}{u-x_B}\left(\frac{1-x_B}{x_B}\right)^{\epsilon/2}
	 v^{\epsilon}w^{\epsilon/2}(1-w)^{\epsilon/2}\,
	\delta(w_r-w)\,dv\,dw\,,
\end{eqnarray}
where now the integration is restricted to the region 
$V=\mbox{B1}\cup\mbox{B2}$. Notice that, for these contributions, the energy
and angle variables, $v$ and $w$, are correlated by the $\delta$ function
which constraints the variables to the surface $w=w_r$. This distinctive 
feature of the two particle phase space already appeared at 
${\cal O}(\alpha_s)$ 
\cite{grau}.


\section{Order $\alpha_s^2$ partonic cross sections}

Order-$\alpha_s^2$ contributions to the one-particle inclusive
cross sections come from the following partonic reactions:
\begin{equation}\label{eq:partreac}
\begin{array}{ll}
\mbox{Real contributions}&\left\{
\begin{array}{lcl}
\gamma+q(\bar{q})&\rightarrow& g+g+q(\bar{q})\\
\gamma+q_{i}(\bar{q_{i}})&\rightarrow&q_{i}(\bar{q_{i}})+q_{j}+\bar{q_{j}}
\,\,\,\,(i\neq j)\\
 \gamma+q_{i}(\bar{q_{i}})&\rightarrow&q_{i}(\bar{q_{i}})+q_{i}+\bar{q_{i}}\\
\gamma+g&\rightarrow& g+q+\bar{q}
\end{array}
\right.\\
\mbox{Virtual contributions}&\left\{
\begin{array}{lcl}
\gamma+q(\bar{q})&\rightarrow& q(\bar{q})\\
\gamma+q(\bar{q})&\rightarrow& g+ q(\bar{q})\\
\gamma+g&\rightarrow& q+\bar{q}
\end{array}
\right.
\end{array}
\end{equation}
where any of the outgoing partons can fragment into the final state hadron 
$h$. Gluon initiated reactions were already discussed in deep in reference 
\cite{oad}. In this section we analyze quark initiated processes, and examine 
the
the nature of the singularities that they involve. These contributions are 
computed in $d=4+\epsilon$ dimensions, in the Feynman gauge, and considering 
all the quarks as massless as in reference \cite{oad}. Algebraic 
manipulations were performed with the aid of the program {\sc Mathematica} 
\cite{math} and the package {\sc Tracer} \cite{tracer} to perform the traces 
over the Dirac indices. 

Taking into account all the alternatives for one parton undergoing
fragmentation into the final state hadron $h$, the reactions in equation
(\ref{eq:partreac}) lead to the following partonic cross sections: 
$\sigma_{qg}$, $\sigma_{q_{i}q_{i}}$, $\sigma_{q_{i}\bar{q_{i}}}$, 
$\sigma_{q_{i}q_{j}}$, $\sigma_{q_{i}\bar{q_{j}}}$, and their charge 
conjugates, exhausting all the possible combinations. The first index 
labels the parton that initiates the reaction, while the second stands 
for the one that precedes the final state hadron. Notice that for the last 
four cross sections, with quarks in both the initial and the hadronized 
state, the are several flavor combinations, each leading to  characteristic
singular contributions.

\begin{figure}[hbt]
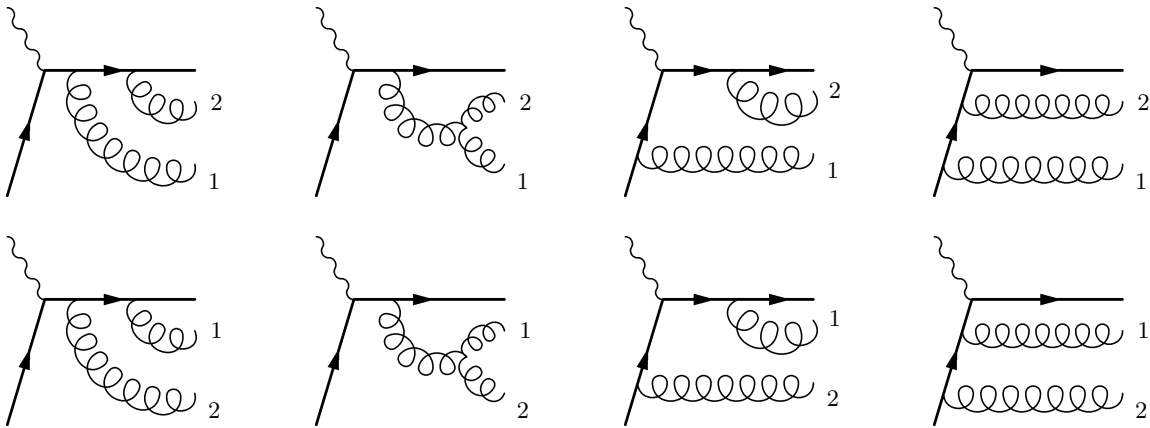

\begin{center}
\begin{minipage}{40mm}
\begin{fmfchar*}(25,25)
  \fmfstraight  \fmfset{arrow_len}{3mm}
  \fmfleft{qi,i1,i2,i3,gamma} 
  \fmfright{g0,g1,g2,g3,qf,o1,o2} 
  \fmf{photon,tension=2,width=5}{gamma,vq}
  \fmf{fermion}{qi,vq}
  \fmf{plain,tension=3}{vq,a1}
  \fmf{fermion,tension=2}{a1,a2}
  \fmf{plain,tension=2}{a2,qf}
  \fmffreeze
  \fmf{gluon,right=.5,width=5}{a1,g1}
  \fmflabel{1}{g1}
  \fmf{gluon,right=.5,width=5}{a2,g3}
  \fmflabel{2}{g3}
\end{fmfchar*}
\end{minipage}
\begin{minipage}{40mm}
\begin{fmfchar*}(25,25)
  \fmfstraight  \fmfset{arrow_len}{3mm}
  \fmfleft{qi,i1,i2,i3,gamma} 
  \fmfright{g0,g1,g2,g3,qf,o1,o2} 
  \fmf{photon,tension=2,width=5}{gamma,vq}
  \fmf{fermion}{qi,vq}
  \fmf{plain,tension=3}{vq,a1}
  \fmf{fermion,tension=2}{a1,a2}
  \fmf{plain,tension=2}{a2,qf}
  \fmffreeze
  \fmf{gluon,left=.8,tension=2,width=5}{gg1,a1}
   \fmf{gluon,right=.2,width=5,tension=3.5}{gg1,g1}
  \fmflabel{1}{g1}
  \fmf{gluon,right=.2,width=5,tension=0.5}{g3,gg1}
  \fmflabel{2}{g3}
\end{fmfchar*}
\end{minipage}
\begin{minipage}{40mm}
\begin{fmfchar*}(25,25)
  \fmfstraight  \fmfset{arrow_len}{3mm}
  \fmfleft{qi,i1,i2,i3,gamma} 
  \fmfright{g0,g1,g1p,,g2,g3,qf,o1,o2,o3} 
  \fmf{photon,tension=2,width=5}{gamma,vq}
  \fmf{phantom,tension=3}{qi,a1,a0,vq}
  \fmf{fermion,tension=0}{qi,vq}
  \fmf{fermion,tension=1.5}{vq,a2,qf}
  \fmffreeze
  \fmf{gluon,width=5}{a1,g1p}
  \fmflabel{1}{g1p}
  \fmf{gluon,right=.5,width=5}{a2,g3}
  \fmflabel{2}{g3}
\end{fmfchar*}
\end{minipage}
\begin{minipage}{40mm}
\begin{fmfchar*}(25,25)
  \fmfstraight  \fmfset{arrow_len}{3mm}
  \fmfleft{qi,i1,i2,i3,gamma} 
  \fmfright{g0,g1,g2,g3,qf,o1,o2} 
  \fmf{photon,tension=2,width=5}{gamma,vq}
  \fmf{phantom,tension=4}{qi,a0,a1,a2,vq}
  \fmf{fermion,tensio=0}{qi,vq}
  \fmf{fermion,tension=0.75}{vq,qf}
  \fmffreeze
  \fmf{gluon,width=5}{a0,g1}
  \fmflabel{1}{g1}
  \fmf{gluon,width=5}{a2,g3}
  \fmflabel{2}{g3}
\end{fmfchar*}
\end{minipage}

\vspace{5mm}

\begin{minipage}{40mm}
\begin{fmfchar*}(25,25)
  \fmfstraight  \fmfset{arrow_len}{3mm}
  \fmfleft{qi,i1,i2,i3,gamma} 
  \fmfright{g0,g1,g2,g3,qf,o1,o2} 
  \fmf{photon,tension=2,width=5}{gamma,vq}
  \fmf{fermion}{qi,vq}
  \fmf{plain,tension=3}{vq,a1}
  \fmf{fermion,tension=2}{a1,a2}
  \fmf{plain,tension=2}{a2,qf}
  \fmffreeze
  \fmf{gluon,right=.5,width=5}{a1,g1}
  \fmflabel{2}{g1}
  \fmf{gluon,right=.5,width=5}{a2,g3}
  \fmflabel{1}{g3}
\end{fmfchar*}
\end{minipage}
\begin{minipage}{40mm}
\begin{fmfchar*}(25,25)
  \fmfstraight  \fmfset{arrow_len}{3mm}
  \fmfleft{qi,i1,i2,i3,gamma} 
  \fmfright{g0,g1,g2,g3,qf,o1,o2} 
  \fmf{photon,tension=2,width=5}{gamma,vq}
  \fmf{fermion}{qi,vq}
  \fmf{plain,tension=3}{vq,a1}
  \fmf{fermion,tension=2}{a1,a2}
  \fmf{plain,tension=2}{a2,qf}
  \fmffreeze
  \fmf{gluon,left=.8,tension=2,width=5}{gg1,a1}
   \fmf{gluon,right=.2,width=5,tension=3.5}{gg1,g1}
  \fmflabel{2}{g1}
  \fmf{gluon,right=.2,width=5,tension=0.5}{g3,gg1}
  \fmflabel{1}{g3}
\end{fmfchar*}
\end{minipage}
\begin{minipage}{40mm}
\begin{fmfchar*}(25,25)
  \fmfstraight  \fmfset{arrow_len}{3mm}
  \fmfleft{qi,i1,i2,i3,gamma} 
  \fmfright{g0,g1,g1p,,g2,g3,qf,o1,o2,o3} 
  \fmf{photon,tension=2,width=5}{gamma,vq}
  \fmf{phantom,tension=3}{qi,a1,a0,vq}
  \fmf{fermion,tension=0}{qi,vq}
  \fmf{fermion,tension=1.5}{vq,a2,qf}
  \fmffreeze
  \fmf{gluon,width=5}{a1,g1p}
  \fmflabel{2}{g1p}
  \fmf{gluon,right=.5,width=5}{a2,g3}
  \fmflabel{1}{g3}
\end{fmfchar*}
\end{minipage}
\begin{minipage}{40mm}
\begin{fmfchar*}(25,25)
  \fmfstraight  \fmfset{arrow_len}{3mm}
  \fmfleft{qi,i1,i2,i3,gamma} 
  \fmfright{g0,g1,g2,g3,qf,o1,o2} 
  \fmf{photon,tension=2,width=5}{gamma,vq}
  \fmf{phantom,tension=4}{qi,a0,a1,a2,vq}
  \fmf{fermion,tensio=0}{qi,vq}
  \fmf{fermion,tension=0.75}{vq,qf}
  \fmffreeze
  \fmf{gluon,width=5}{a0,g1}
  \fmflabel{2}{g1}
  \fmf{gluon,width=5}{a2,g3}
  \fmflabel{1}{g3}
\end{fmfchar*}
\end{minipage}
\end{center}
\caption{Real contributions at order $\alpha_s^2$ for the 
$\gamma+q(\bar{q})\rightarrow g+g+q(\bar{q})$ process}
\label{fig:qgreal}
\end{figure}

\begin{figure}[hbt]
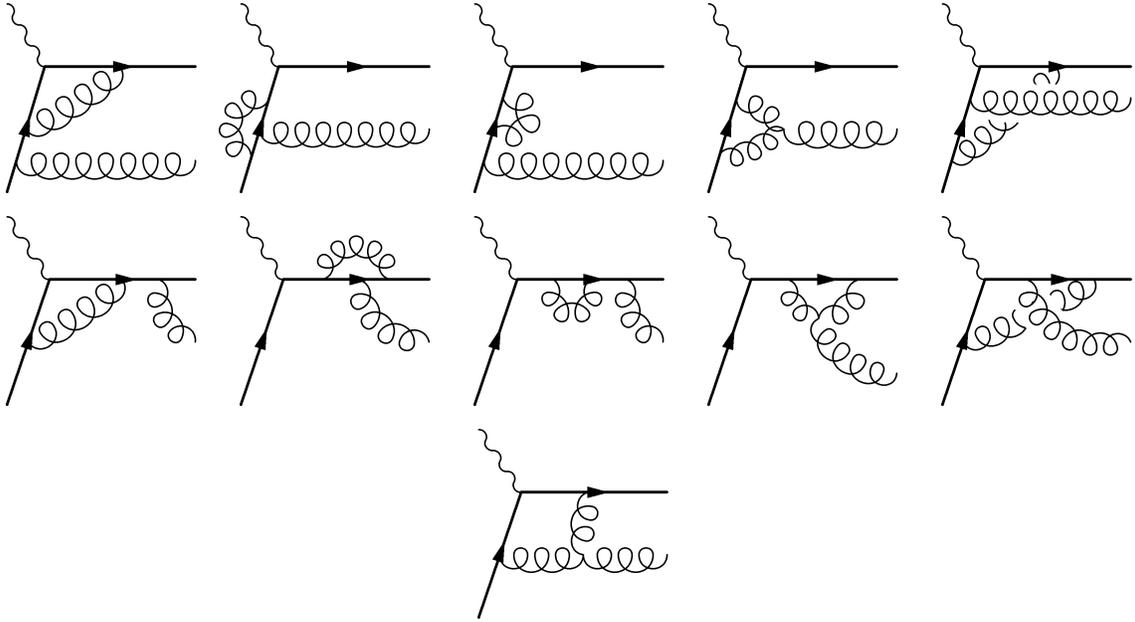

\begin{center}
\begin{minipage}{30mm}
\begin{fmfchar*}(25,25)
  \fmfstraight  \fmfset{arrow_len}{3mm}
  \fmfleft{qi,i1,i2,i3,gamma} 
  \fmfright{g0,g1,g2,g3,qf,o1,o2} 
  \fmf{photon,tension=2,width=5}{gamma,vq}
  \fmf{plain,tension=4}{qi,a0,a1,a2,vq}
  \fmf{fermion,tension=0}{qi,vq}
  \fmf{plain,tension=1.5}{vq,a3,qf}
  \fmf{fermion,tension=0}{vq,qf}
  \fmffreeze
  \fmf{gluon,width=5}{a0,g1}
  \fmf{gluon,width=5}{a1,a3}
\end{fmfchar*}
\end{minipage}
\begin{minipage}{30mm}
\begin{fmfchar*}(25,25)
  \fmfstraight  \fmfset{arrow_len}{3mm}
  \fmfleft{qi,i1,i2,i3,gamma} 
  \fmfright{g0,g1,g2,g3,qf,o1,o2} 
  \fmf{photon,tension=2,width=5}{gamma,vq}
  \fmf{plain,tension=4}{qi,a0,a1,a2,vq}
  \fmf{fermion,tension=0}{qi,vq}
  \fmf{plain,tension=1.5}{vq,a3,qf}
  \fmf{fermion,tension=0}{vq,qf}
  \fmffreeze
  \fmf{gluon,width=5}{a1,g2}
  \fmf{gluon,width=5,left}{a0,a2}
\end{fmfchar*}
\end{minipage}
\begin{minipage}{30mm}
\begin{fmfchar*}(25,25)
  \fmfstraight  \fmfset{arrow_len}{3mm}
  \fmfleft{qi,i1,i2,i3,gamma} 
  \fmfright{g0,g1,g2,g3,qf,o1,o2} 
  \fmf{photon,tension=2,width=5}{gamma,vq}
  \fmf{plain,tension=4}{qi,a0,a1,a2,vq}
  \fmf{fermion,tension=0}{qi,vq}
  \fmf{plain,tension=1.5}{vq,a3,qf}
  \fmf{fermion,tension=0}{vq,qf}
  \fmffreeze
  \fmf{gluon,width=5}{a0,g1}
  \fmf{gluon,width=5,left=2}{a2,a1}
\end{fmfchar*}
\end{minipage}
\begin{minipage}{30mm}
\begin{fmfchar*}(25,25)
  \fmfstraight  \fmfset{arrow_len}{3mm}
  \fmfleft{qi,i1,i2,i3,gamma} 
  \fmfright{g0,g1,g2,g3,qf,o1,o2} 
  \fmf{photon,tension=2,width=5}{gamma,vq}
  \fmf{plain,tension=4}{qi,a0,a1,a2,vq}
  \fmf{fermion,tension=0}{qi,vq}
  \fmf{plain,tension=1.5}{vq,a3,qf}
  \fmf{fermion,tension=0}{vq,qf}
  \fmffreeze
  \fmf{gluon,width=5,left=0.25}{a2,gi,a0}
  \fmf{gluon,width=5}{gi,g2}
\end{fmfchar*}
\end{minipage}
\begin{minipage}{30mm}
\begin{fmfchar*}(25,25)
  \fmfstraight  \fmfset{arrow_len}{3mm}
  \fmfleft{qi,i1,i2,i3,gamma} 
  \fmfright{g0,g1,g2,g3,qf,o1,o2} 
  \fmf{photon,tension=2,width=5}{gamma,vq}
  \fmf{plain,tension=4}{qi,a0,a1,a2,vq}
  \fmf{fermion,tension=0}{qi,vq}
  \fmf{plain,tension=1.5}{vq,a3,qf}
  \fmf{fermion,tension=0}{vq,qf}
  \fmffreeze
  \fmf{gluon,width=5,rubout=10.}{a2,g3}
  \fmf{gluon,width=5}{a0,a3}
\end{fmfchar*}
\end{minipage}
\vspace{3mm}

\begin{minipage}{30mm}
\begin{fmfchar*}(25,25)
  \fmfstraight  \fmfset{arrow_len}{3mm}
  \fmfleft{qi,i1,i2,i3,gamma} 
  \fmfright{g0,g1,g2,g3,qf,o1,o2} 
  \fmf{photon,tension=2,width=5}{gamma,vq}
  \fmf{plain,tension=2}{qi,a3,vq}
  \fmf{fermion,tension=0}{qi,vq}
  \fmf{plain,tension=3.5}{vq,a0,a1,a2,qf}
  \fmf{fermion,tension=0}{vq,qf}
  \fmffreeze
  \fmf{gluon,width=5,left=0.5}{g2,a2}
  \fmf{gluon,width=5}{a3,a1}
\end{fmfchar*}
\end{minipage}
\begin{minipage}{30mm}
\begin{fmfchar*}(25,25)
  \fmfstraight  \fmfset{arrow_len}{3mm}
  \fmfleft{qi,i1,i2,i3,gamma} 
  \fmfright{g0,g1,g2,g3,qf,o1,o2} 
  \fmf{photon,tension=2,width=5}{gamma,vq}
  \fmf{plain,tension=2}{qi,a3,vq}
  \fmf{fermion,tension=0}{qi,vq}
  \fmf{plain,tension=3.5}{vq,a0,a1,a2,qf}
  \fmf{fermion,tension=0}{vq,qf}
  \fmffreeze
  \fmf{gluon,width=5,left=0.5}{g2,a1}
  \fmf{gluon,width=5,left}{a0,a2}
\end{fmfchar*}
\end{minipage}
\begin{minipage}{30mm}
\begin{fmfchar*}(25,25)
  \fmfstraight  \fmfset{arrow_len}{3mm}
  \fmfleft{qi,i1,i2,i3,gamma} 
  \fmfright{g0,g1,g2,g3,qf,o1,o2} 
  \fmf{photon,tension=2,width=5}{gamma,vq}
  \fmf{plain,tension=2}{qi,a3,vq}
  \fmf{fermion,tension=0}{qi,vq}
  \fmf{plain,tension=3.5}{vq,a0,a1,a2,qf}
  \fmf{fermion,tension=0}{vq,qf}
  \fmffreeze
  \fmf{gluon,width=5,left=0.5}{g2,a2}
  \fmf{gluon,width=5,left=2}{a1,a0}
\end{fmfchar*}
\end{minipage}
\begin{minipage}{30mm}
\begin{fmfchar*}(25,25)
  \fmfstraight  \fmfset{arrow_len}{3mm}
  \fmfleft{qi,i1,i2,i3,gamma} 
  \fmfright{g0,g1,g2,g3,qf,o1,o2} 
  \fmf{photon,tension=2,width=5}{gamma,vq}
  \fmf{plain,tension=2}{qi,a3,vq}
  \fmf{fermion,tension=0}{qi,vq}
  \fmf{plain,tension=3.5}{vq,a0,a1,a2,qf}
  \fmf{fermion,tension=0}{vq,qf}
  \fmffreeze
  \fmf{gluon,width=5,left=.5}{a2,gi1,a0}
  \fmf{gluon,width=5,right=0.25}{gi1,g1}
  \fmf{phantom}{a3,gi1}
\end{fmfchar*}
\end{minipage}
\begin{minipage}{30mm}
\begin{fmfchar*}(25,25)
  \fmfstraight  \fmfset{arrow_len}{3mm}
  \fmfleft{qi,i1,i2,i3,gamma} 
  \fmfright{g0,g1,g2,g3,qf,o1,o2} 
  \fmf{photon,tension=2,width=5}{gamma,vq}
  \fmf{plain,tension=2}{qi,a3,vq}
  \fmf{fermion,tension=0}{qi,vq}
  \fmf{plain,tension=3.5}{vq,a0,a1,a2,qf}
  \fmf{fermion,tension=0}{vq,qf}
  \fmffreeze
  \fmf{gluon,width=5,left=0.5,rubout=10.}{g2,a0}
  \fmf{gluon,width=5}{a3,a2}
\end{fmfchar*}
\end{minipage}
\vspace{3mm}

\begin{minipage}{30mm}
\begin{fmfchar*}(25,25)
  \fmfstraight  \fmfset{arrow_len}{3mm}
  \fmfleft{qi,i1,i2,i3,gamma} 
  \fmfright{g0,g1,g2,g3,qf,o1,o2} 
  \fmf{photon,tension=2,width=5}{gamma,vq}
  \fmf{plain,tension=2}{qi,a3,vq}
  \fmf{fermion,tension=0}{qi,vq}
  \fmf{plain,tension=3.5}{vq,a0,a1,a2,qf}
  \fmf{fermion,tension=0}{vq,qf}
  \fmffreeze
  \fmf{gluon,width=5}{a3,gi}
  \fmf{gluon,width=5,tension=0}{a1,gi}
  \fmf{gluon,width=5}{gi,g2}
\end{fmfchar*}
\end{minipage}

\end{center}
\caption{Virtual contributions at order $\alpha_s^2$ for the 
$\gamma+q(\bar{q})\rightarrow g+q(\bar{q})$ process}
\label{fig:qgvirtual}
\end{figure}

Contributions to the $\sigma_{qg}$ cross sections come from the 
first and the sixth reactions in equation (\ref{eq:partreac}), corresponding to
real and virtual diagrams respectively. These are shown in figures 
\ref{fig:qgreal} and \ref{fig:qgvirtual}. Virtual contributions are obtained
from the interference of diagrams in figure \ref{fig:qgvirtual} with those
of single bremsstrahlung at order $\alpha_s$. The diagrams in the second row 
of Figure \ref{fig:qgreal} need to be taken into account due to the symmetric 
character of the outgoing gluons wave function which also contributes 
with a factor $1/2$ to the squared amplitude.

\setlength{\unitlength}{1.mm}

\setlength{\unitlength}{1.mm}

\begin{figure}[hbt]
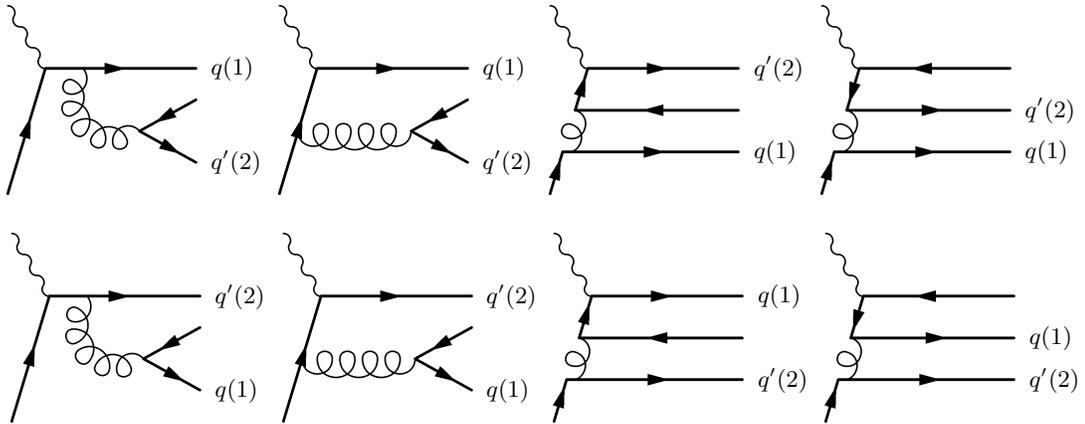

\begin{center}
\begin{minipage}{35mm}
\begin{fmfchar*}(25,25)
  \fmfstraight  \fmfset{arrow_len}{3mm}
  \fmfleft{qi,i1,i2,i3,gamma} 
  \fmfright{g0,g1,g2,g3,qf,o1,o2} 
  \fmf{photon,tension=2,width=5}{gamma,vq}
  \fmf{plain,tension=4}{qi,b1,b2,b3,vq}
  \fmf{fermion,tension=0}{qi,vq}
  \fmf{plain,tension=3}{vq,a1}
  \fmf{fermion,tension=2}{a1,a2}
  \fmf{plain,tension=2}{a2,qf}
  \fmffreeze
  \fmf{phantom}{b2,gi}
  \fmf{gluon,left,width=5,tension=0}{gi,a1}
  \fmf{quark}{g3,gi}
  \fmf{quark}{gi,g1}
  \fmfv{label=$q^{\prime}$(2),label.angle=0}{g1}
  \fmfv{label=$q$(1),label.angle=0}{qf}
\end{fmfchar*}
\end{minipage}
\begin{minipage}{35mm}
\begin{fmfchar*}(25,25)
  \fmfstraight  \fmfset{arrow_len}{3mm}
  \fmfleft{qi,i1,i2,i3,gamma} 
  \fmfright{g0,g1,g2,g3,qf,o1,o2} 
  \fmf{photon,tension=2,width=5}{gamma,vq}
  \fmf{plain,tension=4}{qi,b1,b2,b3,vq}
  \fmf{fermion,tension=0}{qi,vq}
  \fmf{plain,tension=3}{vq,a1}
  \fmf{fermion,tension=2}{a1,a2}
  \fmf{plain,tension=2}{a2,qf}
  \fmffreeze
  \fmf{gluon,width=5}{b2,gi}
  \fmf{quark}{g3,gi}
  \fmf{quark}{gi,g1}
  \fmfv{label=$q^{\prime}$(2),label.angle=0}{g1}
  \fmfv{label=$q$(1),label.angle=0}{qf}
\end{fmfchar*}
\end{minipage}
\begin{minipage}{35mm}
\begin{fmfchar*}(25,25)
  \fmfstraight  \fmfset{arrow_len}{3mm}
  \fmfleft{qi,i1,i2,i3,gamma} 
  \fmfright{g0,g1,g2,g3,g4,g5,qf,o1,o2,o3} 
  \fmf{photon,tension=2,width=5}{gamma,vq}
  \fmf{phantom,tension=6}{qi,b1,b2,b3,b4,b5,vq}
  \fmf{fermion,tension=0}{qi,b2}
  \fmf{fermion,tension=0}{b4,vq}
  \fmf{gluon,tension=0,width=5}{b2,b4}
  \fmf{plain,tension=3}{vq,a1}
  \fmf{fermion,tension=2}{a1,a2}
  \fmf{plain,tension=2}{a2,qf}
  \fmffreeze
  \fmf{quark}{b2,g2}
  \fmf{quark}{g4,b4}
  \fmfv{label=$q^{\prime}$(2),label.angle=0}{qf}
  \fmfv{label=$q$(1),label.angle=0}{g2}
\end{fmfchar*}
\end{minipage}
\begin{minipage}{35mm}
\begin{fmfchar*}(25,25)
  \fmfstraight  \fmfset{arrow_len}{3mm}
  \fmfleft{qi,i1,i2,i3,gamma} 
  \fmfright{g0,g1,g2,g3,g4,g5,qf,o1,o2,o3} 
  \fmf{photon,tension=2,width=5}{gamma,vq}
  \fmf{phantom,tension=6}{qi,b1,b2,b3,b4,b5,vq}
  \fmf{fermion,tension=0}{qi,b2}
  \fmf{fermion,tension=0}{vq,b4}
  \fmf{gluon,tension=0,width=5}{b2,b4}
  \fmf{plain,tension=3}{vq,a1}
  \fmf{fermion,tension=2}{a2,a1}
  \fmf{plain,tension=2}{a2,qf}
  \fmffreeze
  \fmf{quark}{b2,g2}
  \fmf{quark}{b4,g4}
  \fmfv{label=$q^{\prime}$(2),label.angle=0}{g4}
  \fmfv{label=$q$(1),label.angle=0}{g2}
\end{fmfchar*}
\end{minipage}
\vspace{5mm}

\begin{minipage}{35mm}
\begin{fmfchar*}(25,25)
  \fmfstraight  \fmfset{arrow_len}{3mm}
  \fmfleft{qi,i1,i2,i3,gamma} 
  \fmfright{g0,g1,g2,g3,qf,o1,o2} 
  \fmf{photon,tension=2,width=5}{gamma,vq}
  \fmf{plain,tension=4}{qi,b1,b2,b3,vq}
  \fmf{fermion,tension=0}{qi,vq}
  \fmf{plain,tension=3}{vq,a1}
  \fmf{fermion,tension=2}{a1,a2}
  \fmf{plain,tension=2}{a2,qf}
  \fmffreeze
  \fmf{phantom}{b2,gi}
  \fmf{gluon,left,width=5,tension=0}{gi,a1}
  \fmf{quark}{g3,gi}
  \fmf{quark}{gi,g1}
  \fmfv{label=$q^{\prime}$(2),label.angle=0}{qf}
  \fmfv{label=$q$(1),label.angle=0}{g1}
\end{fmfchar*}
\end{minipage}
\begin{minipage}{35mm}
\begin{fmfchar*}(25,25)
  \fmfstraight  \fmfset{arrow_len}{3mm}
  \fmfleft{qi,i1,i2,i3,gamma} 
  \fmfright{g0,g1,g2,g3,qf,o1,o2} 
  \fmf{photon,tension=2,width=5}{gamma,vq}
  \fmf{plain,tension=4}{qi,b1,b2,b3,vq}
  \fmf{fermion,tension=0}{qi,vq}
  \fmf{plain,tension=3}{vq,a1}
  \fmf{fermion,tension=2}{a1,a2}
  \fmf{plain,tension=2}{a2,qf}
  \fmffreeze
  \fmf{gluon,width=5}{b2,gi}
  \fmf{quark}{g3,gi}
  \fmf{quark}{gi,g1}
  \fmfv{label=$q^{\prime}$(2),label.angle=0}{qf}
  \fmfv{label=$q$(1),label.angle=0}{g1}
\end{fmfchar*}
\end{minipage}
\begin{minipage}{35mm}
\begin{fmfchar*}(25,25)
  \fmfstraight  \fmfset{arrow_len}{3mm}
  \fmfleft{qi,i1,i2,i3,gamma} 
  \fmfright{g0,g1,g2,g3,g4,g5,qf,o1,o2,o3} 
  \fmf{photon,tension=2,width=5}{gamma,vq}
  \fmf{phantom,tension=6}{qi,b1,b2,b3,b4,b5,vq}
  \fmf{fermion,tension=0}{qi,b2}
  \fmf{fermion,tension=0}{b4,vq}
  \fmf{gluon,tension=0,width=5}{b2,b4}
  \fmf{plain,tension=3}{vq,a1}
  \fmf{fermion,tension=2}{a1,a2}
  \fmf{plain,tension=2}{a2,qf}
  \fmffreeze
  \fmf{quark}{b2,g2}
  \fmf{quark}{g4,b4}
  \fmfv{label=$q^{\prime}$(2),label.angle=0}{g2}
  \fmfv{label=$q$(1),label.angle=0}{qf}
\end{fmfchar*}
\end{minipage}
\begin{minipage}{35mm}
\begin{fmfchar*}(25,25)
  \fmfstraight  \fmfset{arrow_len}{3mm}
  \fmfleft{qi,i1,i2,i3,gamma} 
  \fmfright{g0,g1,g2,g3,g4,g5,qf,o1,o2,o3} 
  \fmf{photon,tension=2,width=5}{gamma,vq}
  \fmf{phantom,tension=6}{qi,b1,b2,b3,b4,b5,vq}
  \fmf{fermion,tension=0}{qi,b2}
  \fmf{fermion,tension=0}{vq,b4}
  \fmf{gluon,tension=0,width=5}{b2,b4}
  \fmf{plain,tension=3}{vq,a1}
  \fmf{fermion,tension=2}{a2,a1}
  \fmf{plain,tension=2}{a2,qf}
  \fmffreeze
  \fmf{quark}{b2,g2}
  \fmf{quark}{b4,g4}
  \fmfv{label=$q^{\prime}$(2),label.angle=0}{g2}
  \fmfv{label=$q$(1),label.angle=0}{g4}
\end{fmfchar*}
\end{minipage}
\end{center}
\caption{Real contributions at order $\alpha_s^2$ for the 
$\gamma+q(\bar{q})\rightarrow q^{\prime}+\bar{q^{\prime}}+q(\bar{q})$ 
process. Diagrams in the second row only contribute in the case 
$q=q^{\prime}$}
\label{fig:qqreal}
\end{figure}

For the remaining contributions, where a final state quark hadronizes,
it is necessary to discriminate whereas this last quark carries the same 
flavor of the initial state quark or not. Contributions to the 
$\sigma_{q_{i}q_{j}}$ and $\sigma_{q_{i}\bar{q}_{j}}$ cross sections 
with $i\neq j$, arise form the diagrams in the first row of Figure 
\ref{fig:qqreal} corresponding to the second reaction in equation 
(\ref{eq:partreac}). The amplitudes of the 
first two diagrams in this row are proportional to $e_{i}$, whereas 
the two last amplitudes are proportional to $e_{j}$. Consequently, 
the above mentioned cross sections can be decomposed into three pieces, 
proportional to $e_{i}^2$, $e_{j}^2$ and $e_i\,e_j$ respectively. 
Since these cross sections first appear at order-$\alpha_s^2$  
they receive no virtual contributions at this order. 
As the amplitudes are symmetric under the interchange of the quark-antiquark 
pair of flavor $j$, the relation $\sigma_{q_{i}q_{j}}=\sigma_{q_{i}
\bar{q}_{j}}$ is satisfied.

\setlength{\unitlength}{1.mm}

\begin{figure}[h]
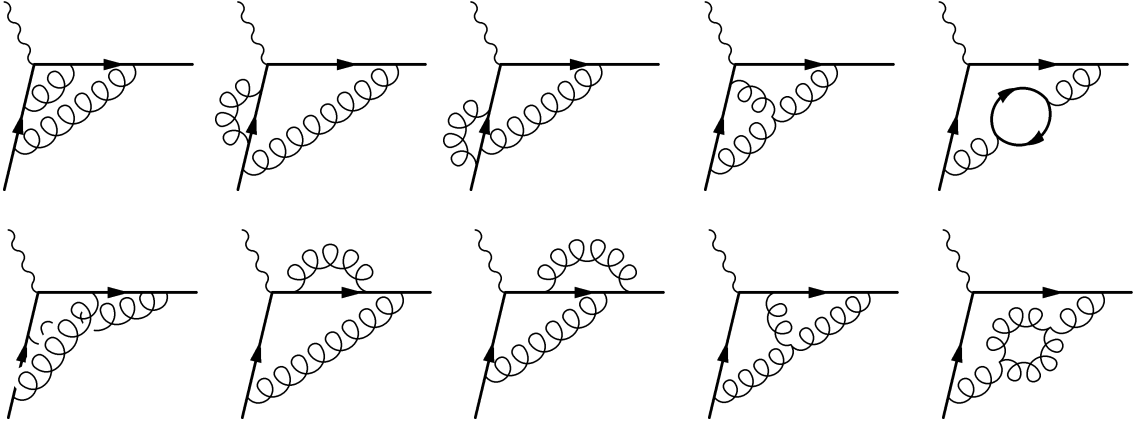

\begin{center}
\begin{minipage}{30mm}
\begin{fmfchar*}(25,25)
  \fmfstraight  \fmfset{arrow_len}{3mm}
  \fmfleft{qi,i1,i2,i3,gamma} 
  \fmfright{g0,g1,g2,g3,qf,o1,o2} 
  \fmf{photon,tension=2,width=5}{gamma,vq}
  \fmf{phantom,tension=6}{qi,b1,b2,b3,b4,b5,vq}
  \fmf{fermion,tension=0}{qi,vq}
  \fmf{plain,tension=5}{vq,a1,a2,a3}
  \fmf{fermion,tension=2}{a3,a4}
  \fmf{plain,tension=3}{a4,a5,qf}
  \fmffreeze
  \fmf{gluon,width=5}{b2,a4}
  \fmf{gluon,width=5}{b4,a2}
\end{fmfchar*}
\end{minipage}
\begin{minipage}{30mm}
\begin{fmfchar*}(25,25)
  \fmfstraight  \fmfset{arrow_len}{3mm}
  \fmfleft{qi,i1,i2,i3,gamma} 
  \fmfright{g0,g1,g2,g3,qf,o1,o2} 
  \fmf{photon,tension=2,width=5}{gamma,vq}
  \fmf{phantom,tension=6}{qi,b1,b2,b3,b4,b5,vq}
  \fmf{fermion,tension=0}{qi,vq}
  \fmf{plain,tension=5}{vq,a1,a2,a3}
  \fmf{fermion,tension=2}{a3,a4}
  \fmf{plain,tension=3}{a4,a5,qf}
  \fmffreeze
  \fmf{gluon,width=5}{b1,a5}
  \fmf{gluon,width=5,left}{b2,b5}
\end{fmfchar*}
\end{minipage}
\begin{minipage}{30mm}
\begin{fmfchar*}(25,25)
  \fmfstraight  \fmfset{arrow_len}{3mm}
  \fmfleft{qi,i1,i2,i3,gamma} 
  \fmfright{g0,g1,g2,g3,qf,o1,o2} 
  \fmf{photon,tension=2,width=5}{gamma,vq}
  \fmf{phantom,tension=6}{qi,b1,b2,b3,b4,b5,vq}
  \fmf{fermion,tension=0}{qi,vq}
  \fmf{plain,tension=5}{vq,a1,a2,a3}
  \fmf{fermion,tension=2}{a3,a4}
  \fmf{plain,tension=3}{a4,a5,qf}
  \fmffreeze
  \fmf{gluon,width=5}{b2,a4}
  \fmf{gluon,width=5,left}{b1,b4}
\end{fmfchar*}
\end{minipage}
\begin{minipage}{30mm}
\begin{fmfchar*}(25,25)
  \fmfstraight  \fmfset{arrow_len}{3mm}
  \fmfleft{qi,i1,i2,i3,gamma} 
  \fmfright{g0,g1,g2,g3,qf,o1,o2} 
  \fmf{photon,tension=2,width=5}{gamma,vq}
  \fmf{phantom,tension=6}{qi,b1,b2,b3,b4,b5,vq}
  \fmf{fermion,tension=0}{qi,vq}
  \fmf{plain,tension=5}{vq,a1,a2,a3}
  \fmf{fermion,tension=2}{a3,a4}
  \fmf{plain,tension=3}{a4,a5,qf}
  \fmffreeze
  \fmf{gluon,width=5}{b1,gi,a4}
  \fmf{gluon,width=5,tension=0}{gi,b5}
\end{fmfchar*}
\end{minipage}
\begin{minipage}{30mm}
\begin{fmfchar*}(25,25)
  \fmfstraight  \fmfset{arrow_len}{3mm}
  \fmfleft{qi,i1,i2,i3,gamma} 
  \fmfright{g0,g1,g2,g3,qf,o1,o2} 
  \fmf{photon,tension=2,width=5}{gamma,vq}
  \fmf{phantom,tension=6}{qi,b1,b2,b3,b4,b5,vq}
  \fmf{fermion,tension=0}{qi,vq}
  \fmf{plain,tension=5}{vq,a1,a2,a3}
  \fmf{fermion,tension=2}{a3,a4}
  \fmf{plain,tension=3}{a4,a5,qf}
  \fmffreeze
  \fmf{gluon,width=5}{b1,gi1}
  \fmf{gluon,width=5}{gi2,a5}
  \fmf{quark,left=0.9,tension=0.5}{gi1,gi2,gi1}
\end{fmfchar*}
\end{minipage}
\vspace{5mm}

\begin{minipage}{30mm}
\begin{fmfchar*}(25,25)
  \fmfstraight  \fmfset{arrow_len}{3mm}
  \fmfleft{qi,i1,i2,i3,gamma} 
  \fmfright{g0,g1,g2,g3,qf,o1,o2} 
  \fmf{photon,tension=2,width=5}{gamma,vq}
  \fmf{phantom,tension=6}{qi,b1,b2,b3,b4,b5,vq}
  \fmf{fermion,tension=0}{qi,vq}
  \fmf{plain,tension=5}{vq,a1,a2,a3}
  \fmf{fermion,tension=2}{a3,a4}
  \fmf{plain,tension=3}{a4,a5,qf}
  \fmffreeze
  \fmf{gluon,width=5,rubout=8}{b1,a3}
  \fmf{gluon,width=5}{b4,a5}
\end{fmfchar*}
\end{minipage}
\begin{minipage}{30mm}
\begin{fmfchar*}(25,25)
  \fmfstraight  \fmfset{arrow_len}{3mm}
  \fmfleft{qi,i1,i2,i3,gamma} 
  \fmfright{g0,g1,g2,g3,qf,o1,o2} 
  \fmf{photon,tension=2,width=5}{gamma,vq}
  \fmf{phantom,tension=6}{qi,b1,b2,b3,b4,b5,vq}
  \fmf{fermion,tension=0}{qi,vq}
  \fmf{plain,tension=5}{vq,a1,a2,a3}
  \fmf{fermion,tension=2}{a3,a4}
  \fmf{plain,tension=3}{a4,a5,qf}
  \fmffreeze
  \fmf{gluon,width=5}{b1,a5}
  \fmf{gluon,width=5,left}{a1,a4}
\end{fmfchar*}
\end{minipage}
\begin{minipage}{30mm}
\begin{fmfchar*}(25,25)
  \fmfstraight  \fmfset{arrow_len}{3mm}
  \fmfleft{qi,i1,i2,i3,gamma} 
  \fmfright{g0,g1,g2,g3,qf,o1,o2} 
  \fmf{photon,tension=2,width=5}{gamma,vq}
  \fmf{phantom,tension=6}{qi,b1,b2,b3,b4,b5,vq}
  \fmf{fermion,tension=0}{qi,vq}
  \fmf{plain,tension=5}{vq,a1,a2,a3}
  \fmf{fermion,tension=2}{a3,a4}
  \fmf{plain,tension=3}{a4,a5,qf}
  \fmffreeze
  \fmf{gluon,width=5}{b2,a4}
  \fmf{gluon,width=5,left}{a2,a5}
\end{fmfchar*}
\end{minipage}
\begin{minipage}{30mm}
\begin{fmfchar*}(25,25)
  \fmfstraight  \fmfset{arrow_len}{3mm}
  \fmfleft{qi,i1,i2,i3,gamma} 
  \fmfright{g0,g1,g2,g3,qf,o1,o2} 
  \fmf{photon,tension=2,width=5}{gamma,vq}
  \fmf{phantom,tension=6}{qi,b1,b2,b3,b4,b5,vq}
  \fmf{fermion,tension=0}{qi,vq}
  \fmf{plain,tension=5}{vq,a1,a2,a3}
  \fmf{fermion,tension=2}{a3,a4}
  \fmf{plain,tension=3}{a4,a5,qf}
  \fmffreeze
  \fmf{gluon,width=5}{b1,gi,a5}
  \fmf{gluon,width=5,tension=0}{a2,gi}
\end{fmfchar*}
\end{minipage}
\begin{minipage}{30mm}
\begin{fmfchar*}(25,25)
  \fmfstraight  \fmfset{arrow_len}{3mm}
  \fmfleft{qi,i1,i2,i3,gamma} 
  \fmfright{g0,g1,g2,g3,qf,o1,o2} 
  \fmf{photon,tension=2,width=5}{gamma,vq}
  \fmf{phantom,tension=6}{qi,b1,b2,b3,b4,b5,vq}
  \fmf{fermion,tension=0}{qi,vq}
  \fmf{plain,tension=5}{vq,a1,a2,a3}
  \fmf{fermion,tension=2}{a3,a4}
  \fmf{plain,tension=3}{a4,a5,qf}
  \fmffreeze
  \fmf{gluon,width=5}{b1,gi1}
  \fmf{gluon,width=5}{gi2,a5}
  \fmf{gluon,left=1,tension=0.5,width=5}{gi1,gi2,gi1}
\end{fmfchar*}
\end{minipage}\end{center}
\caption{Two loops contributions at order $\alpha_s^2$ for the 
$\gamma+q(\bar{q})\rightarrow q(\bar{q})$ 
process (quark form factor).}
\label{fig:ffactor}
\end{figure}

Finally we have the contributions to  $\sigma_{q_{i}q_{i}}$ and 
$\sigma_{q_{i}\bar{q}_{i}}$, where the initial and hadronizing quarks carry the
same flavor. In this case, the relevant reactions are the first three, the
fifth and the sixth in equation (\ref{eq:partreac}) corresponding to diagrams
in Figures \ref{fig:qgreal} and \ref{fig:qqreal} (contributions in both 
$i\neq j$ and $i=j$ cases) for the real contributions and diagrams in 
Figures \ref{fig:qgvirtual} and \ref{fig:ffactor} for the one and two loop 
virtual corrections respectively. Notice, that in spite of having the same 
label for initial and hadronizing quarks, these cross sections include also 
terms proportional to  $e_{i}^2$, $\sum_{j\neq i}e_{j}^2$ and 
$\sum_{j\neq i}e_i\,e_j$, 
since the quark which enters the electromagnetic vertex in the third and 
fourth diagrams in figure \ref{fig:qqreal} can be of any flavor. Regarding 
the process $\gamma+q_{i}\rightarrow q_{i}+q_{i}+\bar{q_{i}}$, diagrams in 
the second row in Figure \ref{fig:qqreal} have a relative minus sign respect to
those in the first row due to the antisymmetry of the wave function of the 
identical outgoing quarks which also contributes with a factor $1/2$ to the 
matrix element. 

\setlength{\unitlength}{1.mm}

The angular integration of the matrix elements for the  real
contributions can be performed with the standard techniques \cite{been}, 
taking into account the additional complications of the one particle 
inclusive case: the necessity of collecting to all orders the potentially 
singular factors in the three particle final state integrals, as explained 
in \cite{oad}. For the integrals that are known to all orders in $\epsilon$, 
this is not a problem, while for those which are only known up to a given 
order a careful treatment is required. In Appendix A we include a list with 
the integrals that appeared in the present calculation.      

Once the angular integrals are performed, matrix elements are still 
distributions in $u$, $v$ and $w$, regulated by the parameter $\epsilon$.
In order to write these contributions, specifically their potentially
singular pieces, in a consistent way, they need to `prescribed', i.e.  
written as an expansion in powers of $\epsilon$ with integrable coefficients.
At variance with the totally inclusive case, where the integration over final 
states leads only to singularities in the $u$ variable, 
in this case they appear along various curves in the residual phase space.
Contributions with singularities along one or two intersecting curves can 
be managed with the technique developed in \cite{oad}. However, processes 
with identical quarks in both initial and hadronizing states lead to 
singularities along three intersecting curves, requiring a different 
prescription strategy. This distinctive feature of the 
$\sigma_{q_{i}{q}_{i}}$ cross section is related to the fact that it give 
rise to quartic poles in $\epsilon$, which cancel out between real and 
virtual contributions.  
 
As an example, we consider the subprocess $\gamma+q\rightarrow q+g+g$, 
where the final state quark fragments into hadron $h$. 
The quark propagators coming from the interference of the first and fourth 
diagrams in the first row of Figure \ref{fig:qgreal} contribute with a factor
\begin{equation}\label{eq:propag}
\frac{1}{(k_1+p_f)^2\,(p_i-k_2)^2\,(k_1+k_2+p_f)^2\,(p_i-k_1-k_2)^2}\,,
\end{equation}
where $p_i$ and $p_f$ are the momenta of the incoming and outgoing quarks 
respectively, while $k_1$ and $k_2$ are those of the gluons. In the center 
of mass frame of the two gluons, the integration over their phase space 
involves only the first two denominators. In this frame, the two other 
denominators do not depend on the angles $\beta_{1}$ and 
$\beta_2$, defined in equation (\ref{eq:PS3}). In this particular case, 
the angular integrals can be performed to all orders in $\epsilon$ and have 
single poles in $\epsilon=0$. The
most singular terms can be written as
\begin{equation}\label{eq:mocopolo}
\frac{1}{\epsilon}\,w^{-1+\epsilon}\,(w-w_{\,0})^{-1+\epsilon/2}\,(w_r-w)^
{\epsilon/2}\,\frac{f(u,v,w,\epsilon)}{1-u}\,,
\end{equation}
where $f(u,v,w)$ is a regular function in all the integration region, 
specifically finite when $\epsilon\rightarrow 0$ and 
$w_{\,0}=-(1-v)/v$. Notice the presence of singularities along $w=0$,
$w=w_{\,0}$ and $u=1$, and that these three curves intersect at the point
$u=v=1$, $w=0$. 

In region B0 only the factor $w^{-1+\epsilon}$ in equation (\ref{eq:mocopolo}) 
is potentially divergent, since $v\le a<1$ and $u\le x_u<1$. This allows
the use of the standard prescription recipe, which lead to a double pole 
accompanied by a $\delta(w)$. In region B1, the singularity at $w=w_{\,0}$
need also to be taken into account, however this can be managed with the 
approach described in \cite{oad}. In this case we obtain
\begin{eqnarray}\label{eq:3B1}
w^{-1+\epsilon}\,(w-w_{\,0})^{-1+\epsilon/2}\,(w_r-w)^{\epsilon/2}
\stackrel{\mbox{B1}}{\longrightarrow}&&\frac{1}{\epsilon}\delta(w)
\,a^{3\epsilon/2}(1-a)^{-\epsilon/2}v^{1-2\epsilon}
\frac{\Gamma(1+\epsilon)\,\Gamma(1+\epsilon/2)}
{\Gamma(1+3\epsilon/2)}\,{}_{2}F_{1}\left[\epsilon,2\epsilon,1+\frac{3}{2}
\epsilon;a\right]\nonumber\\
&&\times\left(\frac{(1-a)^{2\epsilon}}{2\epsilon}\delta(1-v)+
\left((1-v)^{-1+2\epsilon}\right)_{+v[a,\underline{1}]}\right)\nonumber\\
&&+\left(w^{-1+\epsilon}\,(w-w_{\,0})^{-1+\epsilon/2}\,(w_r-w)^{\epsilon/2}
\right)_{+w[\underline{0},w_r]}\,\,.
\end{eqnarray}
Making this substitution in equation (\ref{eq:mocopolo}), we find that, in 
region B1, the leading singularity is a triple pole, proportional to 
$\delta(w)\,\delta(1-v)$. 

Finally, in region B2, the  factor  $(1-u)^{-1}$ is also potentially 
divergent, and should be taken into account. In order to do that, we use 
the prescription given in equation (\ref{eq:3B1}) replacing $a$ with
$z$ as the lower limit in the $v$ prescription, obtaining
\begin{eqnarray}\label{eq:3B2-1}
&&\int_{x_u}^{1}du\,\int_{z}^{1}dv\,\int_{0}^{w_r}dw\,\frac{1}{\epsilon^2}
\delta(w)\,(1-u)^{-1+\epsilon}\left(\frac{(1-z)^{2\epsilon}}{2\epsilon}
\delta(1-v)+\left((1-v)^{-1+2\epsilon}\right)_{+v[z,\underline{1}]}\right)
\tilde{f}(u,v,w,\epsilon)\nonumber\\
&&+\int_{x_u}^{1}du\,\int_{z}^{1}dv\,\int_{0}^{w_r}dw\,\frac{1}{\epsilon}\,
\frac{1}{1-u}\,\left(w^{-1+\epsilon}\,(w-w_{\,0})^{-1+\epsilon/2}
\,(w_r-w)^{\epsilon/2}\right)_{+w[\underline{0},w_r]}\,f(u,v,w,\epsilon)\,,
\end{eqnarray}
In the last equation we have factored out in $\tilde{f}(u,v,w,\epsilon)$ the 
expressions  that are not relevant for the prescription. 
In the first term, the factor $(1-u)^{-1+\epsilon}$ can be prescribed with 
the usual recipe
\begin{equation}
(1-u)^{-1+\epsilon}\rightarrow \frac{1}{\epsilon}\,\delta(1-u)+
\left((1-u)^{-1+\epsilon}\right)_{+u[0,\underline{1}]}\,,
\end{equation}
 whereas the second term can be safely expanded in power series of $\epsilon$,
 since the $u$ integral is convergent even when $\epsilon\rightarrow 0$. 
Indeed, in this case, the pole in $u=1$ is regulated by the fact that the $w$ 
integration region shrinks to the point $w=0$ when $u\rightarrow 1$ whereas
the integrand has no $\delta$ functions. The final result includes
a quartic pole, as mentioned above, accompanied by $\delta(w)\delta(1-v)
\delta(1-u)$. 

All the terms singular along three intersecting curves found in the quark 
initiated corrections can be managed as in the previous example, and they 
present the distinctive feature of having singularities along $u=1$.
Consequently, as it is shown above, the corresponding prescriptions have to be 
modified, but only in region B2.   
Given that the NLO non homogeneous kernels are determined by singularities 
along $w=1$ exclusively, associated to the prescription in the B0 region, 
and to terms which contribute in the forward direction in B1, they receive
no contributions from the more involved triple overlapping singularities.   
 
The virtual one loop contributions to both $\sigma_{qg}$ and $\sigma_{q_{i}
q_{i}}$ can be straightforwardly computed using the standard techniques
\cite{pasa} for handling the loop integrations. These corrections, with 
two partons in the final state, contribute to regions B1 and B2 but not to
region B0. In this case, the singularities relevant for the renormalization
of fracture functions, that is in the $w=1$ direction, are along the curve 
$v=a$ in region B1 and can be managed with the usual prescription recipes in
one variable. On the other side, due to the one-body phase space of the 
outgoing quark, the two loop graphs in Figure \ref{fig:ffactor} only have
support in the point $u=v=1$, $w=0$ and thus their calculation is not 
required for the extraction of the non homogeneous kernels. The results
for these contributions can be found in \cite{KL,matsZP,matsNP}. 
Of course, the complete calculation of the ${\cal O}(\alpha_s^2)$ 
cross sections requires to take into account not only the two loop
quark form factor but also a full analysis of the singularities coming
from the real and one loop virtual contributions, including regions B1 
and B2.


\section{Factorization of singularities and evolution equations}

The order-$\alpha_s^2$ one particle inclusive partonic cross sections 
described in the previous section, have a complex singularity structure, 
revealed as poles in $\epsilon$. Poles associated with ultraviolet 
divergences are removed by coupling constant renormalization. On the other 
side, factorization theorems warrant that poles related to collinear 
singularities 
can be removed from the hadronic cross section by suitable redefinition of 
the bare parton densities, fragmentation and fracture functions in terms of
renormalized ones. The self consistency of these factorization prescriptions 
fixes the scale dependence of the renormalized quantities, and in this case 
allows to obtain the NLO evolution kernels for the fracture functions.

In order to accomplish the factorization procedure, we begin writing the 
hadronic cross sections in terms of renormalized parton densities and 
fragmentation functions. Restricting our attention to the region B0, which 
contain all the forward singularities, we avoid the appearance of divergences 
at $u=1$ or $v=1$ as explained in Section II. 
Notice that in this way, we only have to take into account LO expressions 
for the bare parton densities and fragmentation functions, since NLO 
contributions, once convoluted with the tree level 
partonic cross section, give rise to terms with support in $u=1$ and $v=1$
which contribute only to regions B1 and B2.

As explained in reference \cite{oad}, the convolutions between the LO kernels 
and the ${\cal O}(\alpha_s)$ cross sections can be better handled keeping 
 to all orders in $\epsilon$ the expressions for the cross sections, and 
writing the subtracted 'plus' distributions in the kernels as
\begin{equation}
\left(\frac{1}{1-x}\right)_{+x[0,\underline{1}]}
	\rightarrow \lim_{\epsilon^{\prime}\rightarrow 0}\,\,
	(1-x)^{-1+\epsilon^{\prime}}-
	\frac{1}{\epsilon^{\prime}}\,\delta(1-x)+{\cal O}(\epsilon^{\prime})\,.
\end{equation}
The limit $\epsilon^{\prime}\rightarrow 0$ can be safely taken once the 
convolutions are explicitly evaluated. The resulting counterterms can then be 
subtracted from 
the ${\cal O}(\alpha_s^2)$ cross sections while ultraviolet singularities
are removed by coupling constant renormalization
\begin{equation}
\frac{\alpha_{s}}{2\pi}=\frac{\alpha_s\left(M_R^2\right)}{2\pi}
	\left(1+\frac{\alpha_s\left(M_R^2\right)}{2\pi}\,f_{\Gamma}\,
	\frac{\beta_{0}}{\epsilon}
	\left(\frac{M_R^2}{4\pi\mu^2}\right)^
	{\epsilon/2}\right)\,.
\end{equation}
$\beta_0$ is the lowest order coefficient function in the QCD $\beta$
function,
\begin{equation}
\beta_{0}=\frac{11}{3}\,C_{A}-\frac{4}{3}\,n_f T_F\,
\end{equation}
where $C_A=N$ for $SU(N)$, $T_F=1/2$ as usual, and $n_f$ stands for the 
number of active quark flavors. $M_R$ is the renormalization scale and
\begin{equation}
f_{\Gamma}={\small
	\frac{\Gamma(1+\epsilon/2)}{\Gamma(1+\epsilon)}}
\end{equation}
At this point, only poles proportional to $\delta(1-w)$ are left in region B0,
 to be factorized in the redefinition of the bare 
fracture functions. In terms of renormalized quantities, fracture functions
can be written as
\begin{eqnarray}\label{eq:renfract}
M_{i,h/P}(\xi,\zeta)&=&
	\frac{1}{\xi}\int_{\xi}^{\frac{\xi}{\xi+\zeta}}\frac{du}{u}
	\int_{\frac{\zeta}{\xi}}^{\frac{1-u}{u}}\frac{dv}{v}\,
	\Delta_{ki\leftarrow j}(u,v,M_f)\,f_{j/P}^{r}\left(\frac{\xi}{u},
	M_f^2\right)
	\,D_{h/k}^{r}\left(\frac{\zeta}{\xi\,v},M_f^2\right)\nonumber\\
	&&+\int_{\frac{\xi}{1-\zeta}}^{1}\frac{du}{u}\,
	\Delta_{i\leftarrow j}(u,M_f)\,M_{j,h/P}^{r}
	\left(\frac{\xi}{u},\zeta,M_f^2\right)\,.
\end{eqnarray}
where a sum over repeated indices is understood. Notice that we have chosen 
all 
the factorization scales to be equal to the renormalization scale. 

Given that only single and double 
poles remain in the cross sections once that coupling constant, parton 
densities and fragmentations functions are renormalized, the functions 
$\Delta_{i\leftarrow j}$ and $\Delta_{ki\leftarrow j}$ can be written as
\begin{eqnarray}
\Delta_{i\leftarrow j}(u,M_{f}^2)=\delta_{ij}-\frac{\alpha_s}{2\pi}
	\,f_{\Gamma}\,\left(\frac{M_f^2}{4\pi\mu^2}\right)^{\epsilon/2}
	\frac{2}{\epsilon}\,P^{(0)}_{i\leftarrow j}(u)+
	\left(\frac{\alpha_s}{2\pi}\right)^{2}
	f_{\Gamma}^{\,2}\left(
	\frac{M_f^2}{4\pi\mu^2}\right)^{\epsilon}\Biggl\{
	\frac{1}{\epsilon^2}R^{(1)}_{i\leftarrow j}(u)
	-\frac{1}{\epsilon}P^{(1)}_{i\leftarrow j}(u)
	\Biggr\}  \qquad\,\,\,\, \\
\Delta_{ki\leftarrow j}(u,v,M_{f}^2)=-\frac{\alpha_s}{2\pi}
	\,f_{\Gamma}\,\left(\frac{M_f^2}{4\pi\mu^2}\right)^{\epsilon/2}
	\frac{2}{\epsilon}\,\tilde{P}^{(0)}_{ki\leftarrow j}(u,v)+
	\left(\frac{\alpha_s}{2\pi}\right)^{2}
	f_{\Gamma}^{\,2}\left(
	\frac{M_f^2}{4\pi\mu^2}\right)^{\epsilon}\Biggl\{
	\frac{1}{\epsilon^2}R^{(1)}_{ki\leftarrow j}(u,v) 
 -\frac{1}{\epsilon}
	P^{(1)}_{ki\leftarrow j}(u,v)
	\Biggr\}
\end{eqnarray}
The $\Delta_{i\leftarrow j}$ functions in the homogeneous terms are just those 
found for parton densities in the inclusive case, and can be obtained 
from the corresponding transition functions in reference \cite{zijli}.  The LO 
 non homogeneous kernels $\tilde{P}^{(0)}_{ki\leftarrow j}(u,v)$, can be 
written in terms of those in reference \cite{grau} as
\begin{equation}\label{eq:kertilde}
\tilde{P}^{(0)}_{ki\leftarrow j}(u,v)=P^{(0)}_{ki\leftarrow j}(u)\,
	\delta\left(v-\frac{1-u}{u}\right)\,.
\end{equation}  
Whereas the NLO kernels $P^{(1)}_{ki\leftarrow j}(u,v)$ can only be found
after explicit calculation of the partonic cross sections up to order 
$\alpha_s^2$, the requirement of consistency in the factorization process 
constraints the double pole coefficients $R^{(1)}_{ki\leftarrow j}(u,v)$. 
Indeed, in order to factorization be
consistent, equation (\ref{eq:renfract}) has to be valid no matter the election
of the factorization scale $M_f$ and the corresponding renormalized quantities 
have to be well defined when $\epsilon\rightarrow 0$. These
conditions imply that the ${\cal O}(\alpha_s^2)$ 
coefficients $R^{(1)}_{ki\leftarrow j}(u,v)$ are determined by the 
${\cal O}(\alpha_s)$ kernels. Explicitly,
\begin{eqnarray}\label{eq:Xs}
R^{(1)}_{ki\leftarrow j}(u,v)&=&\sum_{l}\frac{2}{\epsilon^2}\,\left[
\tilde{P}^{(0)}_{ki\leftarrow l}(u,v)\otimes P^{(0)}_{l\leftarrow j}(u)+
\tilde{P}^{(0)}_{li\leftarrow j}(u,v)\otimes P^{(0)}_{k\leftarrow l}(v)+
\tilde{P}^{(0)}_{kl\leftarrow j}(u,v)\otimes^{\prime}P^{(0)}_{i\leftarrow l}(u)
\right]\nonumber\\
&+&\frac{1}{\epsilon^2}\,\beta_{0}\,P^{(0)}_{ki\leftarrow j}(u,v)\,,
\end{eqnarray}
where the convolutions are defined as 
\begin{eqnarray}\label{eq:conv}
f(u,v)\otimes g(u)&=&\int_{u}^{\frac{1}{1+v}}\,\frac{d\bar{u}}{\bar{u}}
	\,f(\bar{u},v)g\left(\frac{u}{\bar{u}}\right)\,,\nonumber\\
f(u,v)\otimes g(v)&=&\int_{v}^{\frac{1-u}{u}}\,\frac{d\bar{v}}{\bar{v}}
	\,f(u,\bar{v})g\left(\frac{v}{\bar{v}}\right)\,,\\
f(u,v)\otimes^{\prime} g(u)&=&\int_{u}^{1-u\,v}\,\frac{d\bar{u}}{\bar{u}}
	\,\frac{u}{\bar{u}}\,f\left(\bar{u},\frac{u}{\bar{u}}v\right)
	g\left(\frac{u}{\bar{u}}\right)\,.\nonumber
\end{eqnarray}
The fact that equation (\ref{eq:Xs}) actually holds for any combination of 
indices is
a useful mean of checking the results and provides a non-trivial verification
of factorization of collinear singularities. 
 
Once the bare fracture functions in the hadronic cross section, equation 
(\ref{eq:hadcsec}), are
written in terms of the renormalized quantities, equation (\ref{eq:renfract}),
the explicit cancellation of forward collinear singularities, showing as single
poles in the partonic cross sections, define the NLO evolution kernels for 
fracture functions. 

Forward singularities coming from the $\sigma_{qg}$ cross section determine
the non homogeneous evolution kernel 
$P^{(1)}_{gq\leftarrow q}(u,v)$. Notice that the first index to the left in 
this notation refers to the parton that fragments into the final state 
hadron, the second labels the flavor of the fracture function that factorizes
the corresponding pole, and the one to the right is for the initial parton.  
Due to charge conjugation symmetry $\sigma_{qg}=\sigma_{\bar{q}g}$ and thus
$P^{(1)}_{gq\leftarrow q}(u,v)=P^{(1)}_{g\bar{q}\leftarrow \bar{q}}(u,v)$. 

The kernels coming from the $\sigma_{q_{i}q_{j}}$ cross sections can be 
classified identifying the cases in which the quark that interacts with the 
photon carries either the flavor of the initial state or that of the quark 
that hadronizes, $P^{(1)}_{q_j q_i\leftarrow q_i}(u,v)$ and 
$P^{(1)}_{q_j\bar{q}_j\leftarrow q_i}(u,v)$ respectively. Notice that in the 
first case the corresponding pole in the cross section will be multiplied by 
$e_{i}^2$, in the second by $e_{j}^2$, while there are no singular 
contributions proportional $e_i\,e_j$. Charge conjugation implies:
\begin{eqnarray}\begin{array}{ccccccc}
P^{(1)}_{q_j q_i\leftarrow q_i} & = &  P^{(1)}_{q_j\bar{q}_i\leftarrow 
\bar{q}_i} & = & P^{(1)}_{\bar{q}_j q_i\leftarrow q_i}
& = & P^{(1)}_{\bar{q}_j \bar{q}_i\leftarrow \bar{q}_i} \\ &&&&&& \\
P^{(1)}_{q_j \bar{q}_j \leftarrow q_i} & = & P^{(1)}_{\bar{q}_j q_j\leftarrow 
q_i} & = & P^{(1)}_{\bar{q}_j q_j\leftarrow \bar{q}_i}
& = & P^{(1)}_{q_j \bar{q}_j\leftarrow \bar{q}_i}
 \end{array} 
\end{eqnarray}  

The cross sections $\sigma_{q_iq_i}$, with the same quark flavor in 
both the initial and final state, can also be classified 
according to the charge factor that accompanies the simple pole. For the case 
in which the charge factor is $e_{j}^2$ with 
$j\neq i$ (that is contributions coming from the first two diagrams in 
Figure \ref{fig:qqreal}), the factorization of forward singularities 
fixes the sum of $P^{(1)}_{q_iq_j\leftarrow q_i} (u,v)$ and 
$P^{(1)}_{q_i\bar{q}_j\leftarrow q_i} (u,v)$, but can not determine them 
individually. However, the symmetry under the exchange of the quark-antiquark 
pair of flavor $j$ of the cross section, implies
\begin{eqnarray}\begin{array}{ccc}
P^{(1)}_{q_iq_j\leftarrow q_i}  & = & 
P^{(1)}_{q_i\bar{q}_j\leftarrow \bar{q}_i}\,,
\end{array}
\end{eqnarray}
whereas charge conjugation gives
\begin{eqnarray}\begin{array}{ccc}
P^{(1)}_{\bar{q}_iq_j\leftarrow q_i}  & = & 
P^{(1)}_{q_iq_j\leftarrow \bar{q}_i}\,.
\end{array}
\end{eqnarray}
Again there are no singularities in the $e_{i}e_{j}$ pieces. 

The terms proportional to $e_{i}^2$ in the $\sigma_{q_iq_i}$ cross section 
contribute to the combination of the $P^{(1)}_{qq\leftarrow q}$ and 
$P^{(1)}_{q\bar{q}\leftarrow q}$ kernels.
In this case, although  charge conjugation implies 
\begin{eqnarray}\begin{array}{ccc}
P^{(1)}_{qq\leftarrow q}  & = & 
P^{(1)}_{\bar{q}\bar{q}\leftarrow \bar{q}}\\
P^{(1)}_{q\bar{q}\leftarrow q}  & = & 
P^{(1)}_{\bar{q}q\leftarrow \bar{q}}\,,
\end{array}
\end{eqnarray}
there is no symmetry at the cross section level 
relating $P^{(1)}_{qq\leftarrow q}$ and 
$P^{(1)}_{q\bar{q}\leftarrow q}$, and thus allowing to disentangle them. 
This means that the cancellation of the forward singularities in this 
particular process would not allow to obtain the scale dependence of 
the valence non-singlet combinations of quark fracture functions. 

On the other hand, singlet combinations of fracture functions, and also those 
non-singlet combinations that include $M_{q,h/P}+M_{\bar{q},h/P}$, evolve with 
precisely the above mentioned combination of kernels. Notice that these last 
two combinations are the ones that actually occur in the hadronic cross 
sections, thus, although the kernels obtained so far
are not enough to perform the evolution of individual flavors, they suffice
to obtain the scale dependence of the cross section. 

Finally, the simple pole in the $\sigma_{q\bar{q}}$ cross section is 
factorized into the $P^{(1)}_{\bar{q}q\leftarrow q}$ kernel, which due to 
charge conjugation satisfies:
\begin{eqnarray}\begin{array}{ccc}
P^{(1)}_{\bar{q}q\leftarrow q}  & = & 
P^{(1)}_{q\bar{q}\leftarrow \bar{q}}\,.
\end{array}
\end{eqnarray}
Explicit expressions for the above mentioned kernels are given in Appendix B.

Having computed the relevant kernels, the evolution equations for the
corresponding fracture function flavor combinations can be obtained 
taking moments in two variables of their 
respective factorization prescriptions, equation (\ref{eq:renfract}), 
as it is done in reference \cite{oad}. Taking the
derivative with respect to the scale $M^2$, the resulting algebraic equations 
for the fracture functions moments, can then be solved using the already 
known explicit solutions for those of the parton densities and fragmentation 
functions. Inverting
moments and using eqs.(17-20) we finally arrive to: 
\begin{eqnarray}\label{eq:fractev}
&&\frac{\partial \,M^{r}_{i,h/P}(\xi,\zeta,M^2)}{\partial \log M^2}=
	\frac{\alpha_s(M^2)}{2\pi}
	\int_{\frac{\xi}{1-\zeta}}^{1}\frac{du}{u}\,
	\left[
	\,P^{(0)}_{i\leftarrow j}(u)
	+\frac{\alpha_s(M^2)}{2\pi} 
	\,P^{(1)}_{i\leftarrow j}(u)\right]
	\,M_{j,h/P}^{r}
	\left(\frac{\xi}{u},\zeta,M^2\right)\nonumber\\
	&&+\frac{\alpha_s(M^2)}{2\pi}\,\frac{1}{\xi}
	\int_{\xi}^{\frac{\xi}{\xi+\zeta}}\frac{du}{u}
	\int_{\frac{\zeta}{\xi}}^{\frac{1-u}{u}}\frac{dv}{v}\,
	\left[
	\tilde{P}^{(0)}_{ki\leftarrow j}(u,v)+\frac{\alpha_s(M^2)}{2\pi}\,
	P^{(1)}_{ki\leftarrow j}(u,v)\right]
	\,f_{j/P}^{r}\left(\frac{\xi}{u},
	M^2\right)
	\,D_{h/k}^{r}\left(\frac{\zeta}{\xi\,v},M^2\right)\,,
\end{eqnarray}
where, again, we have chosen all the scales to be equal. At order 
$\alpha_s$, the Dirac delta of the non homogeneous kernels, 
equation (\ref{eq:kertilde}), allows to trivially perform the integration 
over the $v$ variable for the non homogeneous terms in equation 
(\ref{eq:fractev}),
which can be recasted in the more familiar form of reference \cite{ven}.
However, at NLO the kernels $P^{(1)}_{ki\leftarrow j}(u,v)$ are no
longer constrained to the curve $v=(1-u)/u$ and thus the general 
expressions for the evolution equations involve a double convolution. 


\section{Phenomenology}

In this section we evaluate the phenomenological consequences
of the non homogeneous corrections computed so far, examining the impact of
these corrections relative to the homogeneous evolution \cite{AP}. 
Unfortunately, there
is scarce direct information on semi-inclusive cross sections with a wide
kinematical coverage, as required as input for the evolution equations. 
However there are sensible model estimates available, which can used for the
present purpose, in particular  for $\pi^+$ production from proton targets. 

In order to implement the comparison, we consider a model estimate of the 
corresponding fracture function $M_{q,\pi^+/P}$ based on the ideas 
of reference \cite{Holt} for the description of forward hadrons in DIS, 
already applied to fracture functions in reference\cite{prd97}. In this approach, 
the fracture function is written as
\begin{eqnarray}
M_{q,\pi^{+}/P}(x_B,x_L,Q_0^2),\simeq \phi_{\pi^+/p}(x_L) q(x_B,Q_0^2)\,,
\end{eqnarray}
in terms of the flux of positive pions in the proton times the quark
densities in a neutron. $x_L$ is the ratio between the final state 
hadron energy and that of the proton target in the laboratory frame.
For very forward hadrons, $x_{L}\sim (1-x_{B})\,v_{h}$.
Taking this estimate as an adequate approximation for $M_{q,\pi^+/P}$  
at given initial scale $Q_0^2$, it is possible to obtain the fracture 
densities at higher scales using the evolution equations obtained in the 
last section, and realistic parameterizations for parton densities 
\cite{GRV} and fragmentation functions \cite{kretzer}. 
       
As we have anticipated, although the results obtained in the previous section,
together with those of reference \cite{oad} for the gluon initiated 
contributions, complete the QCD corrections up to order-$\alpha_s^2$ 
in the cross sections and thus the NLO kernels required for $M_q$, 
we still do not know those required for the evolution of the gluon 
fracture density $M_g$. These kernels show up in the order-$\alpha_s^3$ cross 
section and contribute indirectly to the evolution of $M_q$ through
the homogeneous terms in the evolution equations, meanwhile
it is worth assessing the effects of its direct NLO contributions.

In Figure \ref{fig:evxb} we plot the outcome of evolving the model estimate 
for 
$M_{q,\pi^+/P}$ from an initial scale $Q_0^2=1\,GeV^2$ to $Q^2=100\,GeV^2$
as a function of $x_B$ for four different values of $x_L$. The solid
lines correspond to the full NLO evolution for the flavor singlet combination 
of $ M_{q,\pi^{+}/P}$
\begin{equation}
 M_{singlet}^{\pi^+/P}\equiv  \sum_{q,\bar{q}}   M_{q,\pi^{+}/P},
\end{equation}
the dashed lines come from using LO evolution, and the dots show the evolution
using only homogeneous terms.

\setlength{\unitlength}{1.mm}
\begin{figure}[hbt]
\begin{picture}(30,45)(0,0)
\put(-65,-53){\mbox{\epsfxsize7.8cm\epsffile{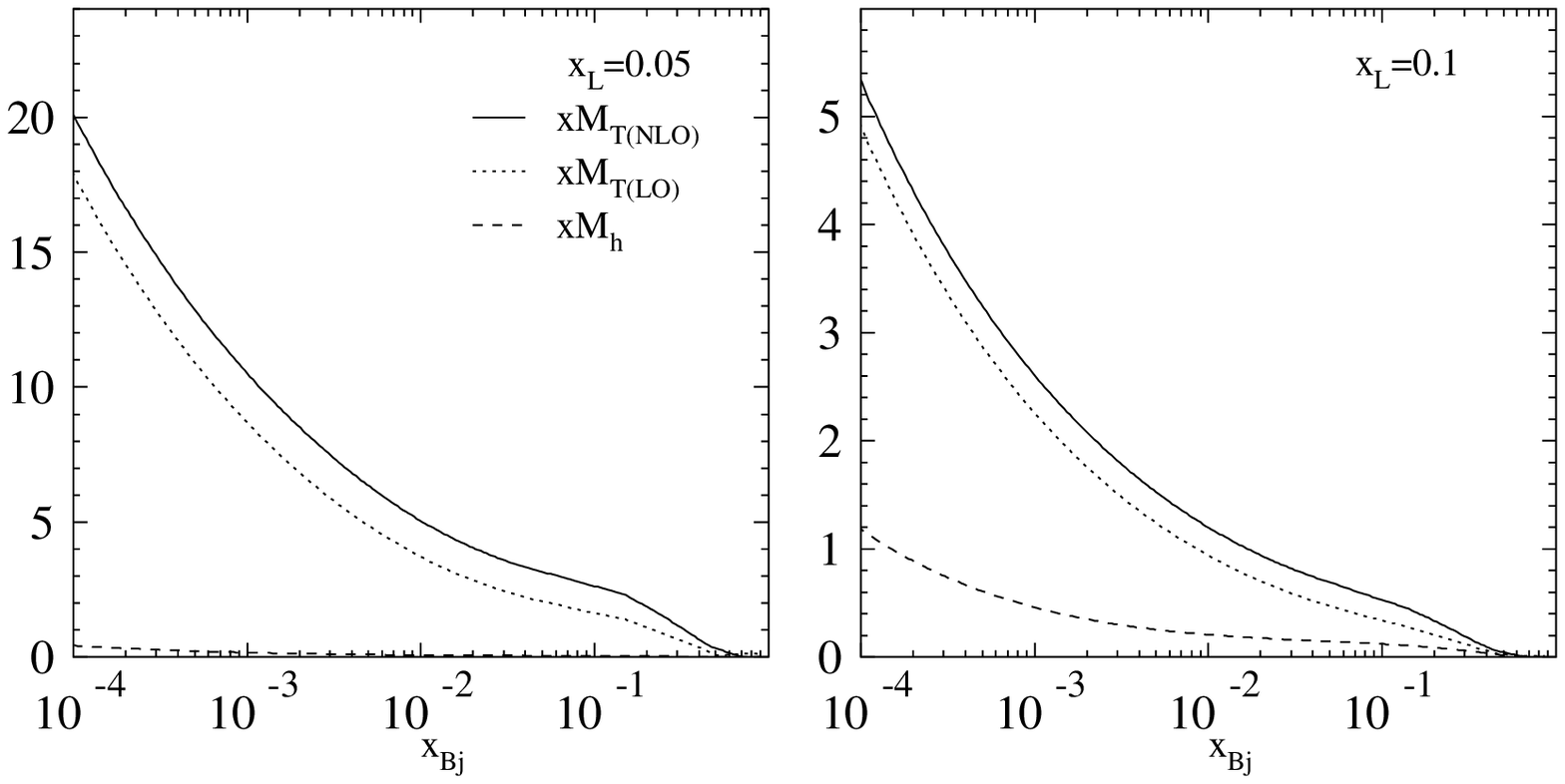}}}
\put(18,-53){\mbox{\epsfxsize7.8cm\epsffile{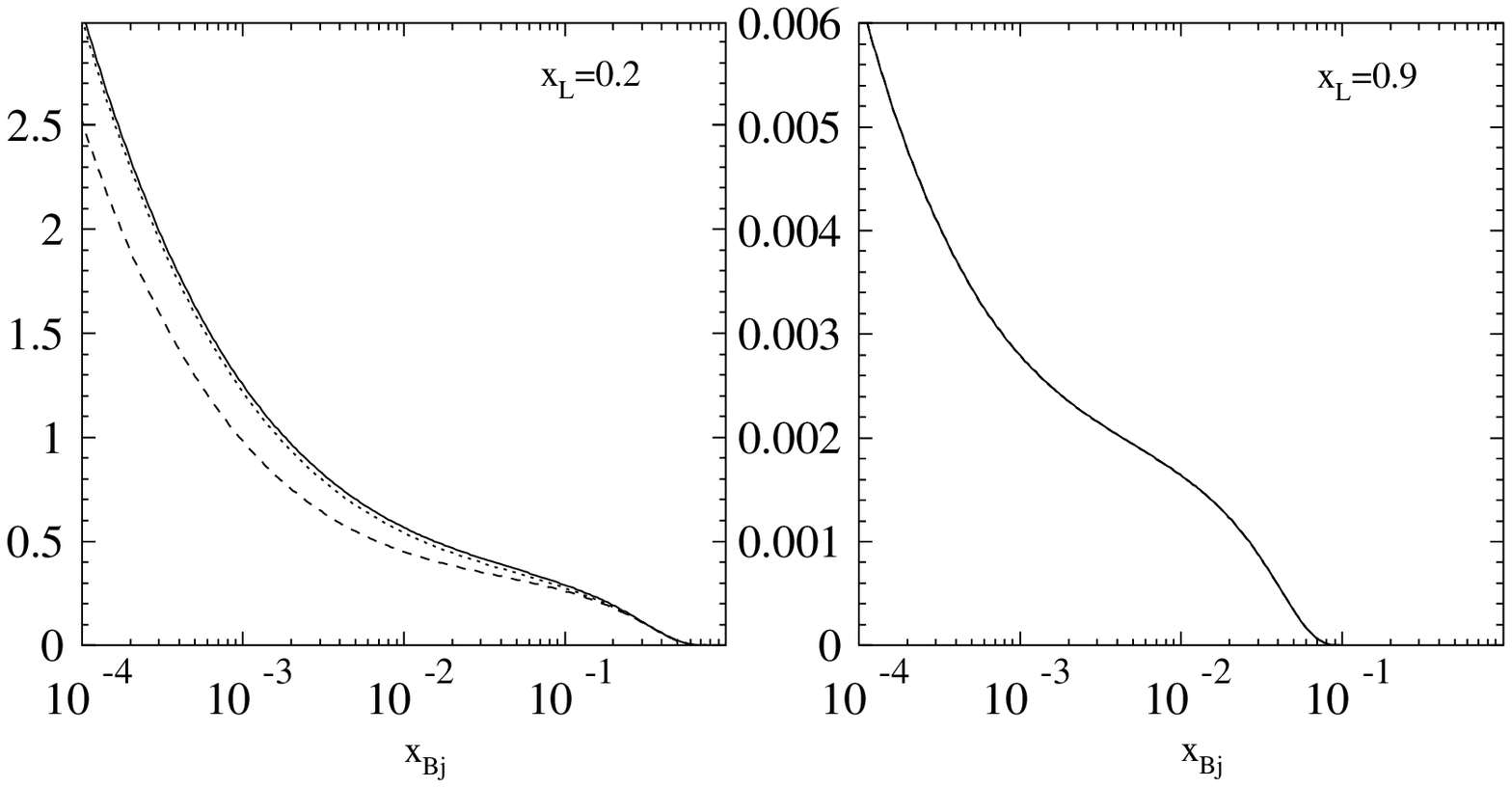}}}
\end{picture}
\caption{Flavor Singlet combination of $x M_q$ at $Q^2=100\, GeV^2$ as evolved
from $Q^2_0=1\,GeV^2$ including  NLO non homogeneus corrections for different
values of the pion energy fraction $x_L$. Solid
lines correspond to the full NLO evolution, dashed lines correspond to LO 
evolution and dotted lines show the evolution
using only homogeneous terms.}\label{fig:evxb}
\end{figure}

NLO non homogeneous corrections clearly have a different relative impact
depending on the kinematical region analysed. For example, in the low 
$x_L$ region, they dominate over the homogeneous terms and their effects 
become as large as those of the LO  terms for large $x_B$. In this region 
the difference between LO and NLO evolution is almost independent of $x_B$, 
however, the LO non homogeneous evolution effects increase considerably
as $x_B$ falls. In Figure \ref{fig:evQ2} we show the same as in Figure 
\ref{fig:evxb} but as a function of 
$Q^2$ for $x_B=0.15$.
 
\setlength{\unitlength}{1.mm}
\begin{figure}[hbt]
\begin{picture}(30,50)(0,0)
\put(-68,-53){\mbox{\epsfxsize7.8cm\epsffile{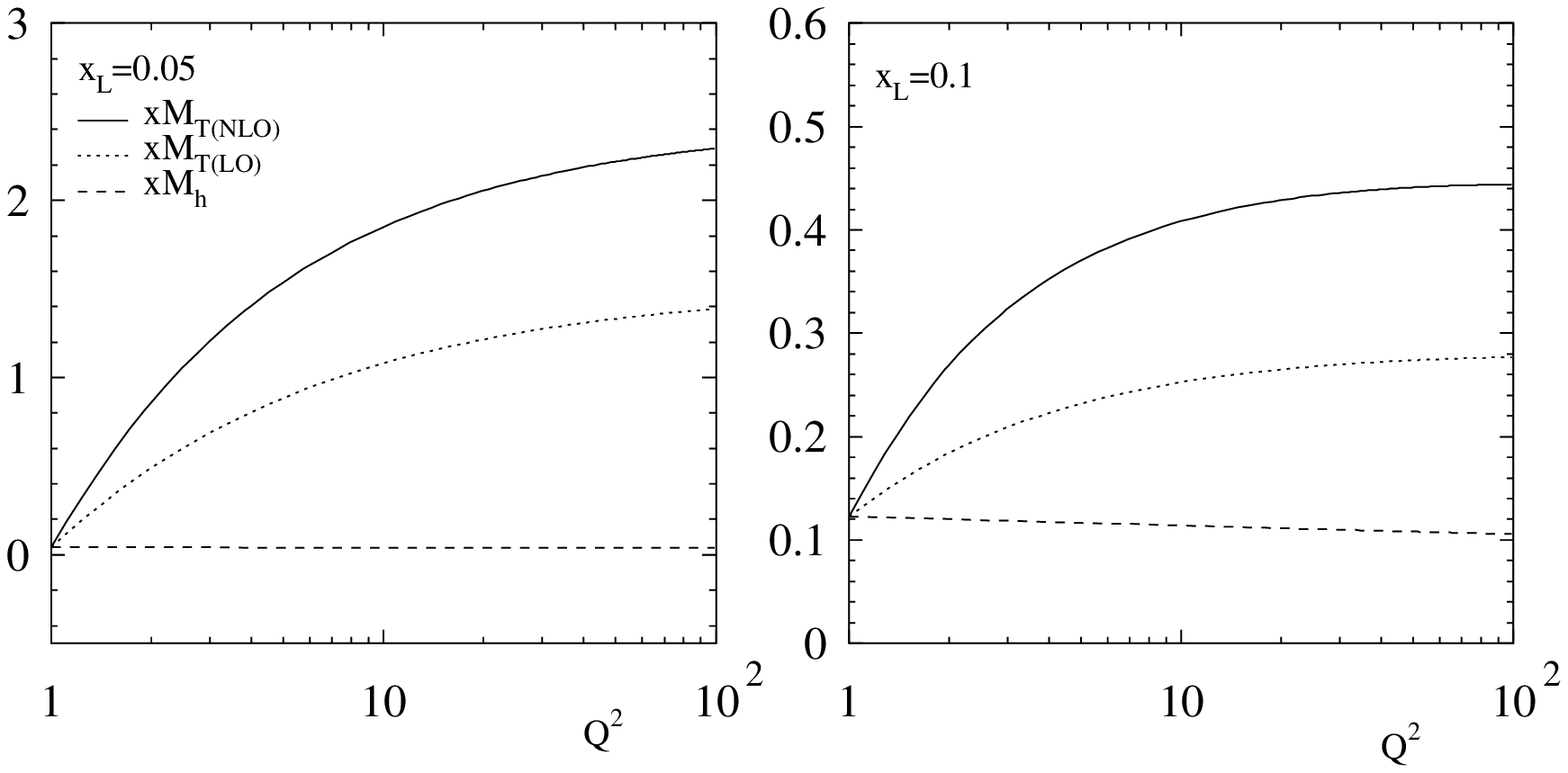}}}
\put(18,-53){\mbox{\epsfxsize7.8cm\epsffile{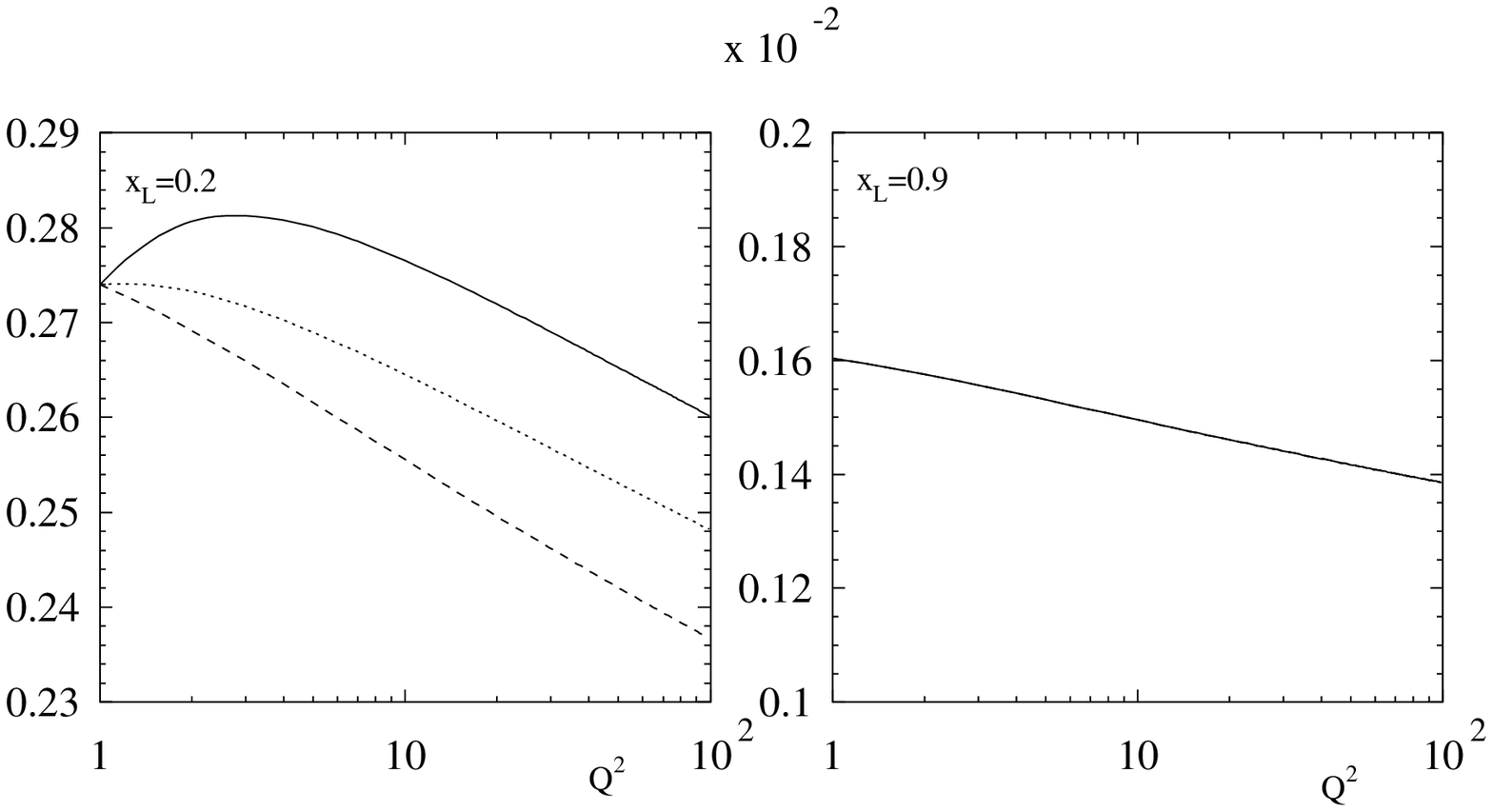}}}
\end{picture}
\caption{Scale dependence of $x M_q$ for $x_B=0.15$}\label{fig:evQ2}
\end{figure}

For larger fractions of hadron momentum, both the LO and NLO non homogeneous 
contributions become more and more suppressed, and although NLO effects
overpower LO corrections, they can not match the growth of the homogeneous
contributions. Large hadron momentum fraction is precisely the region 
relevant for hard diffractive interactions. The lack of significant non 
homogeneus effects in the evolution in this region,  is consistent with 
previous studies on the scale dependence of diffractive deep inelastic 
data \cite{prd98}.

Notice that the relative importance of homogeneous and non 
homogeneous contributions to the evolution depends crucially on the the size
and shape of the appropriate fragmentation and fracture functions corresponding
 to the process considered. This means that the specific energy fraction at 
which non homogeneus effects become non negligible is characteristic of each 
process,and in this particular case, depend on the model estimate used for the
fracture function, as well as on the parameterizations used for parton 
densities and fragmentation functions. From another side, the value chosen  
in the present estimate as departing point for the evolution, 
i.e. the $Q^2$ value   $Q_0^2$ at which we assume the 
model  estimate to hold, modifies to some extent the outcome of the 
evolution and thus the onset of non homogeneous effects. 
In any case, the numerical estimates shown in this section illustrate 
the sort of scale dependence one can expect from a given one particle 
inclusive processes in the light of our present knowledge of QCD corrections.


\section{Conclusions}

 We have computed the ${\cal O}(\alpha_s^2)$ quark initiated corrections to 
one particle inclusive deep inelastic scattering cross sections, verifying and
extending for these processes the feasibility of the approach proposed in 
reference \cite{oad} to deal with overlapping singularities. Although quark 
initiated corrections include a more complex singularity 
structure, once backward configurations are regulated, forward singularities 
can be treated in a similar way to those in gluon initiated corrections. 

We made an explicit check of the factorization of collinear 
singularities arising from quark initiated corrections at this order and 
obtained the evolution kernels required to compute non homogeneous 
contributions to the scale dependence. These kernels, which are labeled by 
three indices, corresponding to the nature of the parton that interacts with 
photon, hadronizes, or initiates the hard processes, respectively, show up 
entangled in the cross section and can not be extracted individually for each
flavor from
SIDIS. However, flavor combinations required for the singlet and total flavor
combinations can be obtained. 
Finally, using a model estimate for fracture functions of protons into pions,
 we found that the impact in the scale dependence of non homogeneous 
NLO corrections relative to LO and to homogeneous terms depends strongly on the
kinematical region under analysis. Specifically, NLO corrections are 
particularly large for small values of parton and pion momentum fractions.

\section{Acknowledgements}

We warmly acknowledge C. A. Garc\'{\i}a Canal, D. de Florian, and G. 
F\'elix for comments and suggestions.


\section*{Appendix A}

In this appendix we present the angular integrals required in the computation
of the quark initiated corrections to SIDIS at order ${\cal O}(\alpha_s^2)$.
As usual, all these integrals can be put, by means of partial 
fractioning, in the standard form \cite{been}
\begin{equation}\label{eq:iang}
I(k,l)=\int_{0}^{\pi}d\beta_1\int_{0}^{\pi}d\beta_2\,
	\frac{\sin^{1+\epsilon}\beta_{1}
	\sin^{\epsilon}\beta_{2}}
	{(a+b\cos\beta_{1})^{k}(A+B\cos\beta_{1}+
	C\sin\beta_{1}\cos\beta_{2})^{l}}\,,
\end{equation}
and can be classified according to whether their parameters 
satisfy either $a^2=b^2$ or $A^2=B^2+C^2$, both 
relations simultaneously, or neither of them. In the following table, these
cases are labeled as $I_a(k,l)$, $I_A(k,l)$, $I_{Aa}(k,l)$ and $I_0(k,l)$
respectively. The parameters $a$, $b$, $A$, $B$ and $C$, are 
functions of the variables $x_B$, $u$, $v$ and $w$. 
The resulting integrals as well as their coefficients can develop 
further singularities associated to these variables, what requires
some care when expanding in power series of the regulator
$\epsilon$. Specifically, the limit $w\rightarrow 1$ implies always that 
$b\rightarrow \pm a$, $C\rightarrow 0$ and $B\rightarrow\pm A$,
which means that factors like $(A\,b-a\,B)^\epsilon$ or 
$(a^2-b^2)^\epsilon$ often present
in the expressions have to be kept to all orders.  This factorization
is trivial for the integrals that can be computed to all orders in 
$\epsilon$, but as it was mentioned in Section II, some special care is 
required for the integrals that have to be expanded before calculation. 
All the integrals of this last type can be handled with the method described 
in reference \cite{oad}.

The integrals that can be done to all orders in 
$\epsilon$ are:
\begin{itemize}
\item $I_{0}$: $a^2\neq b^2$ and $A^2\neq B^2+C^2$.
\begin{align}
&I_0(0,0)=\frac{2\pi}{1+\epsilon}\,,\\
&I_{0}(-1,0)=a\,I_{0}(0,0)\,,\label{eq:i0-10}\\
&I_{0}(-1,-1)=\left[a\,A+\frac{b\,B}{3+\epsilon}\right]\,I_{0}(0,0)\,,\\
&I_{0}(-1,-2)=\left[a\,A^2+\frac{2\,A\,B\,b+a\,(B^2+C^2)}{3+\epsilon}\right]
I_{0}(0,0)\,,\\
&I_{0}(-2,0)=\left[a^2+\frac{b^2}{3+\epsilon}\right]\,I_{0}(0,0)\,,\\
&I_{0}(-2,-1)=\left[a^2\,A+\frac{A\,b^2+2\,a\,b\,B}{3+\epsilon}\right]\,
I_{0}(0,0)\,,\\
&I_{0}(-2,-2)=\left[a^2\,A^2+\frac{A^2\,b^2+4\,A\,a\,B\,b+a^2\,(B^2+C^2)}
{3+\epsilon}+\frac{b^2\,(3\,B^2+C^2)}{(3+\epsilon)(5+\epsilon)}\right]\,
I_{0}(0,0)\,,\\
&I_{0}(-3,0)=\left[a^3+\frac{3\,a\,b^2}{3+\epsilon}\right]\,I_{0}(0,0)\,,\\
&I_{0}(-3,-1)=\left[a^3\,A+\frac{3\,a\,b\,(A\,b+a\,B)}{3+\epsilon}+
\frac{3\,b^3\,B}{(3+\epsilon)(5+\epsilon)}\right]\,I_{0}(0,0)\,,
\label{eq:i0-3-1}\\
&I_{0}(1,0)=\frac{2\pi}{b\,\epsilon}\left\{1-\left(a+b\right)^{-\epsilon}
\left(a^2-b^2\right)^{\epsilon/2}\,{}_{2}F_{1}\left[\frac{\epsilon}{2},
\epsilon,1+\epsilon;\frac{2b}{a+b}\right]\right\}\,,\label{eq:i010}\\
&I_{0}(1,-1)=\frac{A\,b-a\,B}{b}\,I_{0}(1,0)+\frac{2\pi\,B}{(1+\epsilon)\,b}
\,,\\
&I_{0}(1,-2)=\frac{1}{b^2}\left[
(A\,b-a\,B)^2-\frac{C^2(a^2-b^2)}{2+\epsilon}\right]\,I_{0}(1,0)
+\frac{2\pi}{(1+\epsilon)\,b^2}\left[
2A\,B\,b-a\,B^2+\frac{a\,C^2}{2+\epsilon}\right]\,,\\
&I_{0}(1,-3)=\frac{1}{b^3}\left[(A\,b-a\,B)^3-\frac{3\,(A\,b-a\,B)(a^2-b^2)
\,C^2}{2+\epsilon}\right]\,I_{0}(1,0)\notag\\
&\qquad\qquad+\frac{2\pi}{(1+\epsilon)\,b^3}\left[
B\,(3\,A\,b\,(A\,b-a\,B)+a^2 B^2)+\frac{b^2\,B^3}{3+\epsilon}+3\,C^2
\,\left(\frac{B\,b^2}{3+\epsilon}+\frac{a\,(A\,b-a\,B)}{2+\epsilon}
\right)
\right]\,,\\
&I_{0}(1,-4)=\frac{1}{b^4}\left[(A\,b-a\,B)^4-\frac{6\,(a\,b-a\,B)^2(a^2-b^2)
\,C^2}{2+\epsilon}+\frac{3\,(a^2-b^2)^2\,C^2}{(2+\epsilon)(4+\epsilon)}\right]
I_{0}(1,0)\notag\\
&\qquad\qquad+\frac{2\pi}{(1+\epsilon)\,b^4}\Bigg[4\,A\,B\,b\,(A\,b-a\,B)^2
+a\,B^2\,(2\,A^2\,b^2-a^2\,B^2)+\frac{(4\,A\,B\,b-a\,B^3)\,b^2\,B}{3+\epsilon}
\notag\\
&\qquad\qquad+\frac{6\,C^2\,b^2\,B\,(A\,b-a\,B)}{3+\epsilon}+
\frac{6\,C^2\,a\,(A\,b-a\,B)^2}{2+\epsilon}-\frac{3\,a\,C^4\,(a^2-b^2}
{(2+\epsilon)(3+\epsilon)}+\frac{3\,a\,b\,C^2}{(3+\epsilon)(4+\epsilon)}
\Bigg]\,,\\
&I_{0}(2,0)=-\frac{a\,\epsilon}{a^2-b^2}\,I_{0}(1,0)+\frac{2\pi}{a^2-b^2}
\label{eq:i020}\,,\\
&I_{0}(2,-1)=\frac{1}{\epsilon\,a\,b}\left[a\,(A\,b-a\,B)\,\epsilon-
B\,(a^2-b^2)\right]I_{0}(2,0)+\frac{2\pi\,B}{\epsilon\,a\,b}\,,\\
&I_{0}(2,-2)=\frac{1}{\epsilon\,a\,b^2}\left[a\,(A\,b-a\,B)^2\epsilon
-2\,B\,(a^2-b^2)\,
(A\,b-a\,B)-a\,C^2\,(a^2-b^2)\right]\,I_{0}(2,0)\notag\\
&\qquad\qquad+
\frac{2\pi}{\epsilon\,a\,b^2}\left[B\,(2\,A\,b-a\,B)+\frac{a\,(C^2-B^2)}
{1+\epsilon}\right]\,,\\
&I_{0}(2,-3)=\frac{1}{\epsilon\,a\,b^3}\Bigg[a\,(A\,b-a\,B)^3\epsilon-3\,B\,
(a^2-b^2)\,(A\,b-a\,B)^2-3\,C^2\,(a^2-b^2)\notag\\
&\qquad\qquad\times\left(a\,(A\,b-a\,B)-
\frac{B\,(a^2-b^2)}{2+\epsilon}\right)
\Bigg]\,I_{0}(2,0)
+\frac{2\pi}{\epsilon\,a\,b^3}\Bigg[3\,C^2\,\left(\frac{b^2\,B}
{2+\epsilon}+\frac{a\,(A\,b-a\,B)}{1+\epsilon}-\frac{a^2\,B}{(1+\epsilon)
(2+\epsilon)}\right)\notag\\
&\qquad\qquad+B\,\left(3\,A\,b\,(A\,b-a\,B)+a^2\,B^2+\frac{2\,a^2\,B^2
-3\,a\,A\,b\,B}{1+\epsilon}\right)
\Bigg]\,,\label{eq:i02-3}
\end{align}
where $a>b>0$. 
Notice that the cases $a^2=b^2$ and $A^2=B^2+C^2$ corresponding to integrals 
(\ref{eq:i0-10})-(\ref{eq:i0-3-1}) can be easily obtained taking the 
appropriate limit, whereas for integrals (\ref{eq:i010})-(\ref{eq:i02-3})
only the limit $C^2\rightarrow A^2-B^2$ can be taken, giving the corresponding
$I_A$ integral. The hypergeometric function that appears in the cases $k>0$
can be expanded in power series in $\epsilon$ \cite{ingo}:
\begin{align}
{}_{2}F_{1}\left[\frac{\epsilon}{2},
\epsilon,1+\epsilon;z\right]=&1+\frac{\epsilon^2}{2}\,Li_{2}(z)+
\frac{\epsilon^3}{8}\big(\log^2(1-z)\log(z)+
2\,\log(1-z)\,\mbox{Li}_2(1-z)\notag\\
&-2\,\mbox{Li}_3(1-z)+2\,\zeta(3)
-4\,\mbox{Li}_3(z)\big)+{\cal O}(\epsilon^4)
\end{align}

\item $I_{Aa}$: $a^2=b^2$, $A^2-B^2-C^2=0$. Using the results in \cite{solo},
all these integrals can be put in the standard form \cite{ingo}:
\begin{equation}
I_{Aa}(k,l)=\frac{2\pi}{a^k\,A^l}\,a^{-k-l}\frac{\Gamma(1+\epsilon)}
{\Gamma(1+\frac{\epsilon}{2})^2}\,B\left(1-k+\frac{\epsilon}{2},
1-l+\frac{\epsilon}{2}\right)\,{}_{2}F_{1}\left[k,l,1+\frac{\epsilon}{2};
\frac{A-B}{2\,A}\right]\,,
\end{equation}
where $0\le -\frac{C}{A},\frac{B}{A}\le 1$. The hypergeometric functions
appearing in the cases $k,l>0$ should be carefully treated as they usually
develop singularities in the limit $B\rightarrow -A$. Using the relations in
sections (2.8) and (2.9) in \cite{bate} the above mentioned hypergeometric
functions can be written always in terms of
\begin{equation}
{}_{2}F_{1}\left[\frac{\epsilon}{2},\frac{\epsilon}{2},1+\frac{\epsilon}{2};
\frac{A-B}{2\,A}\right]
\end{equation}
which is regular when $B=-A$ and admits the expansion \cite{ingo}:
\begin{align}
{}_{2}F_{1}\left[\frac{\epsilon}{2},
\frac{\epsilon}{2},1+\frac{\epsilon}{2};z\right]=&1+
\frac{\epsilon^2}{4}\,Li_{2}(z)+
\frac{\epsilon^3}{16}\big(\log^2(1-z)\log(z)+
2\,\log(1-z)\,\mbox{Li}_2(1-z)\notag\\
&-2\,\mbox{Li}_3(1-z)+2\,\zeta(3)
-2\,\mbox{Li}_3(z)\big)+{\cal O}(\epsilon^4)
\end{align}
\end{itemize}

All the integrals which cannot be calculated to all orders in $\epsilon$ are
of the classes $I_A(k,l)$ or $I_a(k,l)$ with $k,l>0$. With an adequate choice
of the reference frame, the latter integrals can be recasted
in terms of $I_A(k,l)$ integrals also, which are considerably simpler to 
calculate. The 10 resulting integrals belong to the sets $I_A(1,1)$,
$I_A(2,1)$, $I_A(1,2)$ and $I_A(2,2)$. In each case, the potentially singular
factors have to be extracted, as described in \cite{oad}, before the expansion
in powers of $\epsilon$. With the exception of two integrals of the type
$I_A(1,1)$, all the integrals present potential singularities 
due to a factor $(A\,a-b\,B)^{-n+\epsilon}$ with $n=1,2,3$. In the remaining 
two, corresponding to the $I_A(1,1)$ case, a factor 
$(a^2-b^2)^{\epsilon/2}$ has to be kept to all orders. 
Explicitly:
\begin{align}
&I_{A}(1,1)=\pi\,(A\,a-b\,B)^{-1+\epsilon/2}\,\Bigg\{
\frac{2}{\epsilon}\,\frac{\Gamma(1+\epsilon)}{\Gamma(1+\frac{1}{\epsilon})^2}
\,A^{-\epsilon/2}(a-b)^{\epsilon/2}{}_{2}F_{1}\left[\frac{\epsilon}{2},
\frac{\epsilon}{2},1+\frac{\epsilon}{2};-\frac{b\,(A-B)}{A\,(a-b)}\right]-
\log\left(\frac{A\,(a+b)}{A\,a-b\,B}\right)\notag\\
&\qquad\qquad-\frac{\epsilon}{2}\,\Bigg[\log\left(\frac{a+b}{a-b}\right)\,
\log\left(-\frac{b\,(A-B)}{A\,(a-b)}\right)+\frac{1}{2}\,\log\left(
\frac{A\,(a+b)}{A\,a-b\,B}\right)\,\log\left(\frac{A\,(a+b)}{b^2\,(A-B)^2\,
(A\,a-b\,B)}\right)\notag\\
&\qquad\qquad+\frac{1}{2}\,\mbox{Li}_{2}\left(-\frac{b\,(A+B)}{A\,a-b\,B}
\right)+\mbox{Li}_{2}\left(\frac{A\,a-b\,B}{A\,(a-b)}\right)
\Bigg]+{\cal O}(\epsilon^2)
\Bigg\}\,,\,\,\mbox{case 1}\\
&I_{A}(1,1)=\frac{4\pi}{\epsilon}\,A^{-\epsilon/2}\,(a-b)^{-\epsilon/2}\,
(A\,a-b\,B)^{-1+\epsilon/2}\,{}_{2}F_{1}\left[\frac{\epsilon}{2},
\frac{\epsilon}{2},1+\frac{\epsilon}{2};-\frac{b\,(A-B)}{A\,(a-b)}\right]-
\frac{\pi\,(a^2-b^2)^{\epsilon/2}}{A\,a-b\,B}\,\Bigg\{\frac{2}
{\epsilon}\,A^{\epsilon/2}\notag\\
&\qquad\qquad\times(a-b)^{-\epsilon/2}\,(A\,a-b\,B)^{-\epsilon/2}\,
{}_{2}F_{1}\left[\frac{\epsilon}{2},\epsilon,1+\epsilon;
-\frac{b\,(A+B)}{A\,a-b\,B}\right]-\log\left(
\frac{A\,(a-b)}{A\,a-b\,B}\right)\notag\\
&\qquad\qquad+\frac{\epsilon}{4}\Bigg[\log^2\left(A\,(a-b)\right)-
2\,\log\left(\frac{A}{a-b}\right)\,\log\left(\frac{A\,(a-b)}{A\,a-b\,B}\right)
-\log^2\left(A\,a-b\,B\right)
\Bigg]+{\cal O}(\epsilon^2)
\Bigg\}
\,,\,\,\mbox{case 2}\\
&I_{A}(2,1)=\frac{\pi}{A-B}\,(a+b)^{-1+\epsilon/2}\,\bigg\{
(A+B)\,A^{\epsilon/2}\,(A\,a-b\,B)^{-1-\epsilon/2}\,
{}_{2}F_{1}\left[\frac{\epsilon}{2},\epsilon,1+\epsilon;
-\frac{b\,(A+B)}{A\,a-b\,B}\right]\notag\\
&\qquad\qquad-2\,(a-b)^{-1-\epsilon/2}\,
{}_{2}F_{1}\left[\frac{\epsilon}{2},\epsilon,1+\epsilon;
-\frac{2\,b}{a-b}\right]
\bigg\}+\frac{2\pi}{(a-b)\,(A\,a-b\,B)}+\frac{\pi}{\epsilon}\,\frac
{\Gamma(1+\epsilon)}{\Gamma\left(1+\frac{\epsilon}{2}\right)^2}\,
A^{1-\epsilon/2}\,(a-b)^{-\epsilon/2}\notag\\
&\qquad\qquad\times(A\,a-b\,B)^{-2+\epsilon/2}\,
(2-\epsilon)\,{}_{2}F_{1}\left[\frac{\epsilon}{2},\frac{\epsilon}{2},
1+\frac{\epsilon}{2};-\frac{2\,b}{a-b}\right]\,
{}_{2}F_{1}\left[\frac{\epsilon}{2},\frac{\epsilon}{2},
1+\frac{\epsilon}{2};-\frac{b\,(A-B)}{A\,(a-b)}\right]\notag\\
&\qquad\qquad-
\pi\,(A\,a-b\,B)^{-2+\epsilon/2}\Bigg\{\frac{b\,(A+B)}{a+b}+A\,\log\left(
\frac{A\,(a+b)}{A\,a-b\,B}\right)+\frac{\epsilon}{2}\,\Bigg[
\frac{A\,(a-b)\,(A+B)}{(a+b)\,(A-B)}\,\log(a-b)\notag\\
&\qquad\qquad-\frac{b\,(A+B)}{a+b}\log(A)
-\frac{(A+B)\,(A\,a-b\,B)}{(a+b)\,(A-B)}\log(a+b)+A\,\log\left(
\frac{a+b}{a-b}\right)\,\log\left(-\frac{b\,(A-B)}{A\,(a-b)}\right)\notag\\
&\qquad\qquad-\frac{2\,(A\,(A\,b-a\,B)+(A-B)\,(A\,a-b\,B))}{(a+b)\,(A-B)}\,
\log\left(\frac{A\,(a+b)}{A\,a-b\,B}\right)\notag\\
&\qquad\qquad+\frac{A}{2}\,\log\left(\frac{A\,(a+b)}
{A\,a-b\,B}\right)\,\log\left(\frac{A\,(a+b)}{b^2\,(A-B)^2\,(A\,a-b\,B)}\right)
+A\,\bigg(\mbox{Li}_{2}\left(\frac{A\,a-b\,B}{A\,(a-b)}\right)
+2\,\mbox{Li}_{2}\left(-\frac{b\,(A+B)}{A\,a-b\,B}\right)\notag\\
&\qquad\qquad-\mbox{Li}_{2}\left(-\frac{2\,b}{a-b}\right)
\bigg)
\Bigg]+{\cal O}(\epsilon^2)
\Bigg\}
\,,\\
&I_{A}(1,2)=-\frac{\pi}{(A\,a-b\,B)^2}\Bigg\{\frac{\Gamma(1+\epsilon)}
{4\,\Gamma(1+\frac{1}{\epsilon})^2}\Bigg[(a+b)(2-\epsilon)+\frac{1}{\epsilon}\,
A^{-\epsilon/2}\,(a-b)^{-\epsilon/2}(A\,a-b\,B)^{\epsilon/2}\,\bigg(
2\,(a+b)\,\epsilon-8\,b\notag\\
&\qquad\qquad+\frac{b\,(a-b)\,(A+B)\,(2-\epsilon)\,
(4-\epsilon)}{A\,a-b\,B}\bigg)\,{}_{2}F_{1}\left[\frac{\epsilon}{2},
\frac{\epsilon}{2},1+\frac{\epsilon}{2};-\frac{b\,(A-B)}{A\,(a-b)}\right]
\Bigg]+\frac{1}{A}\,\Bigg[2\,b\,B-A\,(a-b)
\notag\\
&\qquad\qquad+\frac{2\,(A\,a-b\,B)}{2-\epsilon}
-\frac{1}{2}\,b\,(A+B)A^{\epsilon/2}\,(a+b)^{\epsilon/2}
(A\,a-b\,B)^{-\epsilon/2}\,{}_{2}F_{1}\left[\frac{\epsilon}{2},
\epsilon,1+\epsilon;-\frac{b\,(A+B)}{A\,a-b\,B}\right]
\Bigg]\notag\\
&\qquad\qquad+(A\,a-b\,B)^{\epsilon/2}\Bigg[\frac{b^2\,(A^2-B^2)}
{2\,A\,(A\,a-b\,B)}+\frac{b\,(A\,b-a\,B)}{A\,a-b\,B}\log\left(
\frac{A\,(a+b)}{A\,a-b\,B}\right)+\epsilon\bigg[
\frac{A\,(a^2+b^2)-2\,a\,b\,B}{4\,(A\,a-b\,B)}\notag\\
&\qquad\qquad-\frac{b^2\,(A^2-B^2)}{4\,A\,(A\,a-b\,B)}\log\left(A\,(a-b)
\right)-\frac{A\,a+2\,A\,b+B\,b}{4\,A}\log\left(\frac{a+b}{a-b}\right)-
\frac{b\,(A\,b-a\,B)}{A\,a-b\,B}\log\left(\frac{A\,(a+b)}{A\,a-b\,B}\right)
\notag\\
&\qquad\qquad+\frac{A\,(a-b)+2\,B\,b}{4\,A}\log\left(\frac{A\,(a+b)}
{A\,a-b\,B}\right)+\frac{b\,(A\,b-a\,B)}{2\,(A\,a-b\,B)}\log\left(\frac{a+b}
{a-b}\right)\,\log\left(-\frac{b\,(A-B)}{A\,(a-b)}\right)\notag\\
&\qquad\qquad
+\frac{b\,(A\,b-a\,B)}{4\,(A\,a-b\,B)}\,\log\left(\frac{A\,(a+b)}{A\,a-b\,B}
\right)\,\log\left(\frac{A\,(a+b)}{b^2\,(A-B)^2\,(A\,a-b\,B)}\right)
+\frac{b\,(A\,b-a\,B)}{2(A\,a-b\,B)}\,\bigg(\mbox{Li}_{2}\left(-\frac{b\,(A+B)}
{A\,a-b\,B}\right)\notag\\
&\qquad\qquad+\mbox{Li}_{2}\left(\frac{A\,a-b\,B}{A\,(a-b)}\right)
\bigg)
\bigg]+{\cal O}(\epsilon^2)
\Bigg]
\Bigg\}\,,\\
&I_{A}(2,2)=\frac{\pi}{(A\,a-b\,B)^3}\Bigg\{\frac{\Gamma(1+\epsilon)}{8\,
\Gamma\left(1+\frac{\epsilon}{2}\right)^2}\Bigg[-\frac{(2-\epsilon)\,(a+b)}
{a-b}
\,\left(A\,(a-b)\,(4-\epsilon)-2\,(A\,a-b\,B)\right)+\frac{1}{\epsilon}
\,A^{-\epsilon/2}\,(a-b)^{-\epsilon/2}\notag\\
&\qquad\qquad\times(A\,a-b\,B)^{-1+\epsilon/2}\,\bigg(4\,A^2\,b^2\,(3-\epsilon)
\,(4-\epsilon)-2\,\epsilon\,(2-\epsilon)\,(A\,a+b\,B)^2+A\,b\,(A\,(a-b)-B\,
(a+b))\,\epsilon\,(4-\epsilon)^2\notag\\
&\qquad\qquad+2\,A\,a\,B\,b\,(2-\epsilon)\,\epsilon^2-
4B\,b\,(1-\epsilon)\,(4-\epsilon)\,(A\,a\,(2-\epsilon)+B\,b)
\bigg)\,{}_{2}F_{1}\left[\frac{\epsilon}{2},
\frac{\epsilon}{2},1+\frac{\epsilon}{2};-\frac{b\,(A-B)}{A\,(a-b)}\right]
\Bigg]
\notag\\
&\qquad\qquad+\frac{4-\epsilon}{8}\,\Bigg[4\,(A\,(a+b)-3\,b\,(A+B))-\frac
{8\,(A\,a-b\,B)}{2-\epsilon}+2\,b\,(A+B)\,\epsilon+
\frac{b\,(A+B)}{A\,(a+b)}\notag\\
&\qquad\qquad\times A^{\epsilon/2}\,(a+b)^{\epsilon/2}
(A\,a-b\,B)^{-\epsilon/2}\,(2\,A\,(a+b)\,(2-\epsilon)+b\,(A+B)\,\epsilon)
{}_{2}F_{1}\left[\frac{\epsilon}{2},
\epsilon,1+\epsilon;-\frac{b\,(A+B)}{A\,a-b\,B}\right]
\Bigg]\notag\\
&\qquad\qquad+(A\,a-b\,B)^{\epsilon/2}\Bigg[-\frac{b^2\,(A+B)\,
(5\,(A\,b-a\,B)+3\,(A\,a-b\,B)}
{2\,(a+b)\,(A\,a-b\,B)}+\frac{b\,(3\,A\,(A\,b-a\,B)+B\,(A\,a-b\,B))}
{A\,a-b\,B}\notag\\
&\qquad\qquad\times\log\left(
\frac{A\,(a+b)}{A\,a-b\,B}\right)+\epsilon\bigg[
\frac{1}{4}\left(A\,(a-b)-\frac{2\,b^2\,(A-B)}{a-b}+\frac{2\,b^2\,(A+B)^2}
{A\,(a+b)}-\frac{2\,b^2\,(A^2-B^2)}{A\,a-b\,B}
\right)\notag\\
&\qquad\qquad+\frac{b^2\,(A+B)\,(5\,(A\,b-a\,B)+3\,(A\,a-b\,B)}
{4\,(a+b)\,(A\,a-b\,B)}\log\left(A\,(a-b)
\right)-\frac{6\,b\,(A\,b-a\,B)-(a+3\,b)\,(A\,a-b\,B)}{4\,(a-b)}\notag\\
&\qquad\qquad\times
\log\left(\frac{a+b}{a-b}\right)+
\frac{(A\,a-b\,B)\,(A\,a+2\,A\,b+b\,B)-16\,A\,b\,(A\,b-a\,B)}
{4\,(A\,a-b\,B)}\log\left(\frac{A\,(a+b)}{A\,a-b\,B}\right)
\notag\\
&\qquad\qquad+\frac{b\,(A\,a-b\,B+2\,(A\,b-a\,B)}{2\,(a+b)}
\log\left(\frac{A\,(a+b)}{A\,a-b\,B}\right)+
\frac{b\,(3\,A\,(A\,b-a\,B)+B\,(A\,a-b\,B))}
{2\,(A\,a-b\,B)}\log\left(\frac{a+b}
{a-b}\right)\notag\\
&\qquad\qquad\times\log\left(-\frac{b\,(A-B)}{A\,(a-b)}\right)
+\frac{b\,(3\,A\,(A\,b-a\,B)+B\,(A\,a-b\,B))}{4\,(A\,a-b\,B)}
\,\log\left(\frac{A\,(a+b)}{A\,a-b\,B}
\right)\,\log\left(\frac{A\,(a+b)}{b^2\,(A-B)^2\,(A\,a-b\,B)}\right)
\notag\\
&\qquad\qquad+\frac{b\,(3\,A\,(A\,b-a\,B)+B\,(A\,a-b\,B))}{2(A\,a-b\,B)}\,
\bigg(\mbox{Li}_{2}\left(-\frac{b\,(A+B)}
{A\,a-b\,B}\right)
+\mbox{Li}_{2}\left(\frac{A\,a-b\,B}{A\,(a-b)}\right)
\bigg)
\bigg]+{\cal O}(\epsilon^2)
\Bigg]
\Bigg\}\,.\\
\end{align} 
For the $I_{A}(1,1)$ integrals, case 1 means that in the limit $w\rightarrow 1$
the parameters satisfy $b\rightarrow -a$ and $B\rightarrow -A$, whereas in case
2 they satisfy $b\rightarrow -a$ and $B\rightarrow A$. In this last case the
factors $(A\,a-b\,B)^{-n}$ remain finite when $w\rightarrow 1$.

Notice that in the above list, there are some integrals that can be 
obtained  from another by a suitable transformation, for example, integral 
$I_{0}(2,0)$ can be obtained from $I_{0}(1,0)$ by derivation with respect to 
parameter $a$. Further relations can be obtained using the results presented
in \cite{ingo}. 

\section*{Appendix B}

In this appendix we present results for the NLO non homogeneous kernels
for fracture functions obtained from the factorization of collinear
singularities in the ${\cal O}(\alpha_s^2)$ SIDIS cross section.

In the case where a gluon is the fragmenting particle, we obtained
\begin{align}
&P^{(1)}_{gq\leftarrow q}(u,v)\nonumber\\
&\qquad =\,{{{C_F}}^2}\,\Bigg\{- 
     {\frac{3 - u}{1 + v}} + {\frac{2\,{u^2}}
       {{{\left( 1 - u\,v \right) }^2}}} 
+2\,u + {\frac{2}{{{\left( 1 + v \right) }^2}}}
- {\frac{\left( 3 - u \right) \,u}{1 - u\,v}} + 
     \Bigg[{\frac{{u^2}}{{{\left( 1 - u\,v \right) }^2}}} 
-{{\left( 1 + v \right) }^{-2}} - {\frac{1 + u}{1 + v}}\notag\\
&\qquad\qquad\qquad+ {\frac{u\,\left( 1 + u \right) }{1 - u\,v}} \Bigg] \,
      \left( \log (u) + \log (a_{f} - v) - \log (v) \right)  + 
     \Bigg[2\,p_{gq\leftarrow q}(u)\left(\frac{1}{v} 
+ 
        {\frac{u}{1 - u - u\,v}}\right)
-2\,u\notag\\
&\qquad\qquad\qquad- {\frac{2}{{{\left( 1 + v \right) }^2}}} - 
        {\frac{2\,\left( 1 + u \right) }{1 + v}}  \Bigg]
 \left( \log (1 + v) + \log (1 - u\,v) \right)  + 
     \Big[ u - 2\,\left( 1 - u \right) \,\log (u) + 
        2\,p_{gq\leftarrow q}(u)\notag\\
&\qquad\qquad\qquad
\times\big(\zeta(2)- Li_{2}(u) - 2\,\log (1 - u)\,\log (u) \big)  \Big]
\delta (a_{f} - v)\Bigg\} \notag\\
&\qquad+\,{C_A}\,{C_F}\,\Bigg\{ {\frac{2\,u\,
\left( 2 - 3\,u + 2\,{u^2} \right) }
       {{{\left( 1 - u \right) }^2}}} - 
     {\frac{2\,\left( 1 - u \right) }{v}} - 
     {\frac{4\,{u^2}\,\left( 1 + u + {u^2} \right) \,v}
       {{{\left( 1 - u \right) }^3}}} + 
     {\frac{4\,{u^3}\,\left( 1 + u + {u^2} \right) \,{v^2}}
       {{{\left( 1 - u \right) }^4}}}\notag\\
&\qquad\qquad\qquad
+ \Bigg[ {\frac{u\,\left( 11 - 2\,u + 7\,{u^2} \right) }
          {{{\left( 1 - u \right) }^2}}}+ p_{gq\leftarrow q}(u)\,\left( 
           {\frac{4\,{u^3}\,{v^2}}{{{\left( 1 - u \right) }^3}}}
 {-\frac{7}{v}} - 
           {\frac{4\,{u^2}\,v}{{{\left( 1 - u \right) }^2}}}  - 
           {\frac{3\,u}{1 - u - u\,v}} \right)  \Bigg] \,\log (1 - u)\notag \\
&\qquad\qquad\qquad
+ \Bigg[ -{\frac{u\,\left( 9 - 4\,u + 3\,{u^2} \right) }
           {{{\left( 1 - u \right) }^2}}}
+ {\frac{{u^2}\,\left( 3 - 2\,u + 3\,{u^2} \right) \,v}
          {{{\left( 1 - u \right) }^3}}}
 +  p_{gq\leftarrow q}(u)\,\left( {\frac{7}{v}} - 
           {\frac{2\,{u^3}\,{v^2}}{{{\left( 1 - u \right) }^3}}} + 
           {\frac{2\,u}{1 - u - u\,v}} \right)  \Bigg] \notag\\
&\qquad\qquad\qquad
\times\log (u) +\Bigg[ -{\frac{u\,\left( 5 + 3\,{u^2} \right) }
           {{{\left( 1 - u \right) }^2}}} + 
        {\frac{{u^2}\,{{\left( 1 + u \right) }^2}\,v}
          {{{\left( 1 - u \right) }^3}}}
+ p_{gq\leftarrow q}(u)\,\left( {\frac{3}{v}} - 
           {\frac{2\,{u^3}\,{v^2}}{{{\left( 1 - u \right) }^3}}} \right) 
         \Bigg]
\log (a_{f} - v) \notag\\
&\qquad\qquad\qquad
+\Bigg[-{\frac{2\,u\,\left( 3 - u + 2\,{u^2} \right) }
          {{{\left( 1 - u \right) }^2}}} + 
        {\frac{{u^2}\,\left( 3 - 2\,u + 3\,{u^2} \right) \,v}
          {{{\left( 1 - u \right) }^3}}} + 
        p_{gq\leftarrow q}(u)\left( {\frac{4}{v}} - 
           {\frac{2\,{u^3}\,{v^2}}{{{\left( 1 - u \right) }^3}}} + 
           {\frac{3\,u}{1 - u - u\,v}} \right)  \Bigg]\notag\\
&\qquad\qquad\qquad
\times\log (v) + \Bigg[ 
        {\frac{u\,\left( 1 + u + u\,v \right) }{1 - u}} 
-p_{gq\leftarrow q}(u){\frac{u}{1 - u - u\,v}}  \Bigg]\log (1 + v)
+ \Bigg[ 
        {\frac{u\,\left( 2 - u\,v \right) }{1 - u}} 
-{\frac{p_{gq\leftarrow q}(u)}{v}}  \Bigg] \notag\\
&\qquad\qquad\qquad
\times\log (1 - u\,v)      
-\Bigg[ 2\,\left( 1 - u \right)  + 
        4\,p_{gq\leftarrow q}(u)\,\log\left(a_{f}\right) \Bigg] \,
      {\left({{\frac{1}{a_{f} - v}}}\right)_{+v[0,\underline{a_{f}}]}} \notag\\
&\qquad\qquad\qquad+ 
     4\,p_{gq\leftarrow q}(u)\,{\left({{\frac{\log (a_{f} - v)}
           {a_{f} - v}}}\right)_{+v[0,\underline{a_{f}}]}}+ 
\Big[ 2\,p_{gq\leftarrow q}(u)\,
         \left( \mbox{Li}_{2}(u)
+ \log (1 - u)\,\log (u) \right)\notag\\
&\qquad\qquad\qquad-u 
         \Big] \,\delta (a_{f} - v) \Bigg\}\,,
\end{align}
where
\begin{equation}
p_{qg\leftarrow q}(x)=\frac{1+x^2}{1-x}\,,\,\,\mbox{and}\,\,\,
a_f=\frac{1-u}{u}\,.
\end{equation}
For the kernels relating different quark species, comig from the
quark fragmenting processes, we obtained
\begin{align}
\begin{split}
P^{(1)}_{q_j q_i\leftarrow q_i}(u,v)=
{C_F}\,{T_F}\,\Bigg\{&{2\,{u^2}\,\frac{ \left( 1 + u + {u^2} 
\right) \,v - 
          {u^3}\,\left( 1 - v + {v^2} \right)  - 
          \left( 1 - u + {u^2} \right) \,
           \left( 1 + 4\,u\,v + 2\,u\,{v^2} \right)  + 
          2\,p_{q\leftarrow g}(u) }{{{\left( 1 - u \right) }^4}}}\\
&+ p_{gq\leftarrow q}(u)\,\left( 
          \log (a_{f} - v) + \log (v) 
-\log (a_{f}) - \log (1 - u) \right) \\
&\times \frac{u\,
        \left( 1 - 2\,u\,\left( 1 + v \right)  + 
          {u^2}\,p_{q\leftarrow g}(v) \right) }{{{\left( 1 - u \right) }^3}}
       \Bigg\}\,,
\end{split}\\
\begin{split}
P^{(1)}_{q_{j}\bar{q}_{j}\leftarrow q_{i}}(u,v)=
{C_F}\,{T_F}\,
\Bigg\{&
{-2\,\frac{{u^2}\,v\,{{\left( 1 + v \right) }^2} + 
          {{\left( 1 - v \right) }^2}\,
           \left( 1 - u\,\left( 1 + v \right)  \right)}{u\,
        {{\left( 1 + v \right) }^4}}}
+    \big( \log (u) + \log (a_{f} - v) - \log (v) \\
&+ 2\,\log (1 + v) \big)
\frac{\left( 1 + {v^2} \right) \,
        \left( {u^2}\,{{\left( 1 + v \right) }^2} + 
          2\,\left( 1 - u\,\left( 1 + v \right)  \right)  \right) \,
         }{u\,{{\left( 1 + v \right) }^4}}\Bigg\}\,,
\end{split}\\
\begin{split}
P^{(1)}_{q_{i}q_{j}\leftarrow q_{i}}(u,v)=
{C_F}\,{T_F}\,
\Bigg\{&{2\,{u^2}\,\frac{u\,\left( 2 + {u^2} \right) \,
             {v^2}   + 7\,u\,v\,\left( 1 - u - u\,v \right)  - 
          \left( 1 - u \right) \,\left( 1 + v + u\,v + {u^2}\,{v^3} \right) 
           }{{{\left( 1 - u\,v \right) }^4}}}\\
& +\left[ \log (u)+ \log (a_{f} - v) + \log (v) 
 - 2\,\log\left(\frac{1 - u - u\,v}{1-u\,v}\right) 
\right]{\frac{u\,\left( 1 + {u^2}\,{v^2} \right)}{{{\left( 1 - 
            u\,v \right) }^4}}}\\
&\times
 \left( 1 + {u^2}\,{v^2} - 
          2\,\left( 1 - u \right) \,u\,\left( 1 + v \right)  \right)
\Bigg\}\,,
\end{split}
\end{align}
where
\begin{equation}
p_{q\leftarrow g}(x)=1-2x+2x^2\,.
\end{equation}
Finally, for the kernels relating quarks and antiquarks of the same flavor,
\begin{align}
\begin{split}
P^{(1)}_{\bar{q}q\leftarrow q}(u,v)=
{C_F}\,{T_F}
\,\Bigg\{&-\frac{2\,u\,\left( 1 + 2\,u^2 \right) }
     {3\,{\left( 1 - u \right) }^2} + 
    \frac{4\,u^2\,\left( 4 + 3\,u + 5\,u^2 \right) \,v}
     {3\,{\left( 1 - u \right) }^3} - 
    \frac{2\,u^3\,\left( 3 + 2\,u + 3\,u^2 \right) \,v^2}
     {{\left( 1 - u \right) }^4} - \frac{8}{u\,{\left( 1 + v \right) }^4} \\
&+ \frac{8\,\left( 1 + u \right) }{u\,{\left( 1 + v \right) }^3} - 
    \frac{2\,\left( 3 + 10\,u - 3\,u^2 \right) }
     {3\,u\,{\left( 1 + v \right) }^2} + 
    \frac{4\,\left( 1 - 2\,u \right) }{3\,\left( 1 + v \right) } + 
    \Bigg[ -\frac{2\,u\,\left( 5 - 3\,u + 4\,u^2 \right) }
        {3\,{\left( 1 - u \right) }^2} \\
&+ \frac{2\,u^2\,\left( 7 - 2\,u + 7\,u^2 \right) \,v}
        {3\,{\left( 1 - u \right) }^3} - 
       \frac{4\,u^3\,\left( 1 + u^2 \right) \,v^2}
        {{\left( 1 - u \right) }^4} + 
       \frac{2\,\left( 1 + u^2 \right) }
        {3\,\left( 1 - u \right) \,\left( 1 + v \right) } \Bigg] \,
     \log (1 - u)\\
& + \Bigg[ \frac{2\,u\,\left( 5 - 6\,u + 4\,u^2 \right) }
        {3\,{\left( 1 - u \right) }^2} - 
       \frac{4\,u^2\,\left( 2 - u + 2\,u^2 \right) \,v}
        {3\,{\left( 1 - u \right) }^3} + 
       \frac{2\,u^3\,\left( 1 + u^2 \right) \,v^2}
        {{\left( 1 - u \right) }^4} + 
       \frac{4}{u\,{\left( 1 + v \right) }^4}\\
& - \frac{4\,\left( 1 + u \right) }{u\,{\left( 1 + v \right) }^3} + 
       \frac{2\,{\left( 1 + u \right) }^2}
        {u\,{\left( 1 + v \right) }^2} - 
       \frac{4\,\left( 2 - u^2 \right) }
        {3\,\left( 1 - u \right) \,\left( 1 + v \right) } \Bigg] \,
     \log (u) + \Bigg[ \frac{2\,u\,\left( 5 - 6\,u + 4\,u^2 \right) }
        {3\,{\left( 1 - u \right) }^2}\\
& - \frac{4\,u^2\,\left( 2 - u + 2\,u^2 \right) \,v}
        {3\,{\left( 1 - u \right) }^3} + 
       \frac{2\,u^3\,\left( 1 + u^2 \right) \,v^2}
        {{\left( 1 - u \right) }^4} + 
       \frac{4}{u\,{\left( 1 + v \right) }^4} - 
       \frac{4\,\left( 1 + u \right) }{u\,{\left( 1 + v \right) }^3} + 
       \frac{2\,{\left( 1 + u \right) }^2}
        {u\,{\left( 1 + v \right) }^2} \\
&- \frac{4\,\left( 2 - u^2 \right) }
        {3\,\left( 1 - u \right) \,\left( 1 + v \right) } \Bigg] \,
     \log (a_{f} - v) + 
    \Bigg[ \frac{2\,u^2}{{\left( 1 - u \right) }^2} - 
       \frac{2\,u^2\,\left( 1 + u^2 \right) \,v}
        {{\left( 1 - u \right) }^3} + 
       \frac{2\,u^3\,\left( 1 + u^2 \right) \,v^2}
        {{\left( 1 - u \right) }^4} \\
&- \frac{4}{u\,{\left( 1 + v \right) }^4} + 
       \frac{4\,\left( 1 + u \right) }{u\,{\left( 1 + v \right) }^3} - 
       \frac{2\,{\left( 1 + u \right) }^2}
        {u\,{\left( 1 + v \right) }^2} + 
       \frac{2\,\left( 1 + u \right) }{1 + v} \Bigg] \,\log (v) + 
    \Bigg[ \frac{2\,\left( 1 - 2\,u \right) \,u}
        {3\,\left( 1 - u \right) } + 
       \frac{2\,u^2\,v}{3\,\left( 1 - u \right) } \\
&+ \frac{8}{u\,{\left( 1 + v \right) }^4} - 
       \frac{8\,\left( 1 + u \right) }{u\,{\left( 1 + v \right) }^3} + 
       \frac{4\,{\left( 1 + u \right) }^2}{u\,{\left( 1 + v \right) }^2} - 
       \frac{2\,\left( 5 - 7\,u^2 \right) }
        {3\,\left( 1 - u \right) \,\left( 1 + v \right) } \Bigg] \,
     \log (1 + v) \Bigg\}\,,
\end{split}
\end{align}
and
\begin{align}
&P^{(1)}_{qq\leftarrow q}(u,v)+P^{(1)}_{q\bar{q}\leftarrow q}(u,v)\nonumber\\
&\qquad={C_F}\,{T_F}\,\Bigg\{-\frac{2\,u\,\left( 2 + u^2 \right) }
     {3\,{\left( 1 - u \right) }^2} + 
    \frac{4\,u^2\,\left( 5 + 3\,u + 4\,u^2 \right) \,v}
     {3\,{\left( 1 - u \right) }^3} - 
    \frac{2\,u^3\,\left( 3 + 2\,u + 3\,u^2 \right) \,v^2}
     {{\left( 1 - u \right) }^4} - \frac{8}{u\,{\left( 1 + v \right) }^4}
\notag\\
&\qquad\qquad\qquad\quad + \frac{8\,\left( 1 + u \right) }
{u\,{\left( 1 + v \right) }^3} - 
    \frac{2\,\left( 3 + 14\,u - 3\,u^2 \right) }
     {3\,u\,{\left( 1 + v \right) }^2} - 
    \frac{4\,\left( -2 + u \right) }{3\,\left( 1 + v \right) } - 
    \frac{16\,u^3}{{\left( 1 - u\,v \right) }^4} + 
    \frac{16\,u^2\,\left( 1 + u \right) }{{\left( 1 - u\,v \right) }^3}\notag\\
&\qquad\qquad\qquad\quad - \frac{4\,u\,\left( 1 + 4\,u - u^2 \right) }
     {{\left( 1 - u\,v \right) }^2} + 
    \frac{2\,u\,\left( 5 - 12\,u + 6\,u^2 \right) }
     {3\,\left( 1 - u \right) \,\left( 1 - u\,v \right) } + 
    \bigg[- \frac{2\,u\,\left( 4 - u + 3\,u^2 \right) }
        {3\,{\left( 1 - u \right) }^2}\notag\\
&\qquad\qquad\qquad\quad+ \frac{2\,u^2\,\left( 5 + 2\,u + 5\,u^2 \right) \,v}
        {3\,{\left( 1 - u \right) }^3} - 
       \frac{4\,u^3\,\left( 1 + u^2 \right) \,v^2}
        {{\left( 1 - u \right) }^4} + 
       \frac{2\,u\,\left( 1 + u^2 \right) }
        {3\,\left( 1 - u \right) \,\left( 1 - u\,v \right) } \bigg] \,
     \log (1 - u)\notag\\
&\qquad\qquad\qquad\quad+ \bigg[ \frac{2\,u\,\left( 2 + u \right) }
        {3\,{\left( 1 - u \right) }^2} - 
       \frac{2\,u^2\,\left( 1 + 4\,u + u^2 \right) \,v}
        {3\,{\left( 1 - u \right) }^3} + 
       \frac{2\,u^3\,\left( 1 + u^2 \right) \,v^2}
        {{\left( 1 - u \right) }^4} + 
       \frac{4}{u\,{\left( 1 + v \right) }^4} - 
       \frac{4\,\left( 1 + u \right) }{u\,{\left( 1 + v \right) }^3}\notag\\ 
&\qquad\qquad\qquad\quad+ \frac{2\,{\left( 1 + u \right) }^2}
        {u\,{\left( 1 + v \right) }^2}- 
        \frac{2\,\left( 1 + u \right) }{1 + v} - 
       \frac{8\,u^3}{{\left( 1 - u\,v \right) }^4} + 
       \frac{8\,u^2\,\left( 1 + u \right) }
        {{\left( 1 - u\,v \right) }^3} - 
       \frac{4\,u\,{\left( 1 + u \right) }^2}
        {{\left( 1 - u\,v \right) }^2}\notag\\
&\qquad\qquad\qquad\quad + \frac{2\,u\,\left( 5 - 8\,u^2 \right) }
        {3\,\left( 1 - u \right) \,\left( 1 - u\,v \right) } \bigg] \,
     \log (u) + \Bigg[ \frac{2\,u^2}{{\left( 1 - u \right) }^2} - 
       \frac{2\,u^2\,\left( 1 + u^2 \right) \,v}
        {{\left( 1 - u \right) }^3} + 
       \frac{2\,u^3\,\left( 1 + u^2 \right) \,v^2}
        {{\left( 1 - u \right) }^4}\notag\\
&\qquad\qquad\qquad\quad + \frac{4}{u\,{\left( 1 + v \right) }^4} - 
       \frac{4\,\left( 1 + u \right) }{u\,{\left( 1 + v \right) }^3} + 
       \frac{2\,{\left( 1 + u \right) }^2}
        {u\,{\left( 1 + v \right) }^2} - 
       \frac{2\,\left( 1 + u \right) }{1 + v} - 
       \frac{8\,u^3}{{\left( 1 - u\,v \right) }^4} + 
       \frac{8\,u^2\,\left( 1 + u \right) }
        {{\left( 1 - u\,v \right) }^3} \notag\\
&\qquad\qquad\qquad\quad- \frac{4\,u\,{\left( 1 + u \right) }^2}
        {{\left( 1 - u\,v \right) }^2} 
+ \frac{2\,u\,\left( 5 - 6\,u^2 \right) }
        {3\,\left( 1 - u \right) \,\left( 1 - u\,v \right) } \Bigg] \,
     \log (a_{f} - v) + 
    \Bigg[ \frac{2\,u\,\left( 4 - 4\,u + 3\,u^2 \right) }
        {3\,{\left( 1 - u \right) }^2} \notag\\
&\qquad\qquad\qquad\quad- \frac{4\,u^2\,\left( 1 + u + u^2 \right) \,v}
        {3\,{\left( 1 - u \right) }^3} + 
       \frac{2\,u^3\,\left( 1 + u^2 \right) \,v^2}
        {{\left( 1 - u \right) }^4} - 
       \frac{4}{u\,{\left( 1 + v \right) }^4} + 
       \frac{4\,\left( 1 + u \right) }{u\,{\left( 1 + v \right) }^3} - 
       \frac{2\,{\left( 1 + u \right) }^2}
        {u\,{\left( 1 + v \right) }^2} \notag\\
&\qquad\qquad\qquad\quad+ \frac{2\,\left( 1 + u \right) }{1 + v} + 
       \frac{8\,u^3}{{\left( 1 - u\,v \right) }^4} - 
       \frac{8\,u^2\,\left( 1 + u \right) }
        {{\left( 1 - u\,v \right) }^3} + 
       \frac{4\,u\,{\left( 1 + u \right) }^2}
        {{\left( 1 - u\,v \right) }^2} - 
       \frac{2\,u\,\left( 8 - 5\,u^2 \right) }
        {3\,\left( 1 - u \right) \,\left( 1 - u\,v \right) } \Bigg] \,
     \log (v) \notag\\
&\qquad\qquad\qquad\quad+\Bigg[ 2\,u + \frac{8}{u\,{\left( 1 + v \right) }^4} -
        \frac{8\,\left( 1 + u \right) }{u\,{\left( 1 + v \right) }^3} + 
       \frac{4\,{\left( 1 + u \right) }^2}
        {u\,{\left( 1 + v \right) }^2} - 
       \frac{2\,\left( 7 + 12\,u + 7\,u^2 \right) }
        {3\,\left( 1 + u \right) \,\left( 1 + v \right) }\notag\\
&\qquad\qquad\qquad\quad - \frac{2\,u\,\left( 1 + u^2 \right) }
        {3\,\left( 1 + u \right) \,\left( 1 - u\,v \right) } \Bigg] \,
     \log (1 + v) + \Bigg[ \frac{2\,\left( 5 - 6\,u \right) \,u}
        {3\,\left( 1 - u \right) } - 
       \frac{2\,u^2\,v}{3\,\left( 1 - u \right) } - 
       \frac{2\,\left( 1 + u^2 \right) }
        {3\,\left( 1 + u \right) \,\left( 1 + v \right) }\notag\\
&\qquad\qquad\qquad\quad + \frac{16\,u^3}{{\left( 1 - u\,v \right) }^4} - 
       \frac{16\,u^2\,\left( 1 + u \right) }
        {{\left( 1 - u\,v \right) }^3} + 
       \frac{8\,u\,{\left( 1 + u \right) }^2}
        {{\left( 1 - u\,v \right) }^2} + 
       \frac{4\,u^2}{3\,\left( 1 + u \right) \,\left( 1 - u\,v \right) }
\notag\\
&\qquad\qquad\qquad\quad - \frac{28\,u\,\left( 1 + 2\,u + u^2 \right) }
        {3\,\left( 1 + u \right) \,\left( 1 - u\,v \right) } \Bigg] \,
     \log (1 - u\,v) \Bigg\}\,.
\end{align}
As mentioned in Section III, in this last case, only the combination
$P^{(1)}_{qq\leftarrow q}(u,v)+P^{(1)}_{q\bar{q}\leftarrow q}(u,v)$ can be 
extracted.
The remaining quark initiated kernels can be obtained from the charge 
conjugation relations given in that Section. 


\end{fmffile}
\end{document}